\documentclass[citeautoscript,floatfix,superscriptaddress,twocolumn,showpacs,prb,aps,10pt]{revtex4-1}
\usepackage{graphicx} \usepackage{amsmath, amssymb, bm}
\usepackage{dcolumn} \usepackage{color}
\usepackage{eso-pic}
\usepackage{fix-cm}

\usepackage{multirow}
\usepackage{sidecap}

\usepackage{marginnote}
\usepackage{ulem} %package for the strike out \sout and \uwave

\newcommand{\sss}{\scriptscriptstyle}
\newcommand{\sst}{\scriptstyle}

\newcommand{\stext}[1]{\sss \text{#1} \sst}

\renewcommand{\emph}[1]{\textit{#1}}

\begin{document}
\title{Anisotropy and phonon modes from analysis of the dielectric function tensor and inverse dielectric function tensor of monoclinic yttrium orthosilicate}

\author{A. Mock}
\email{amock@huskers.unl.edu}
\homepage{http://ellipsometry.unl.edu}
\affiliation{Department of Electrical and Computer Engineering and Center for Nanohybrid Functional Materials, University of Nebraska-Lincoln, U.S.A.}
\author{R.~Korlacki}
\affiliation{Department of Electrical and Computer Engineering and Center for Nanohybrid Functional Materials, University of Nebraska-Lincoln, U.S.A.}
\author{S.~Knight}
\affiliation{Department of Electrical and Computer Engineering and Center for Nanohybrid Functional Materials, University of Nebraska-Lincoln, U.S.A.}
\author{M. Schubert}
\affiliation{Department of Electrical and Computer Engineering and Center for Nanohybrid Functional Materials, University of Nebraska-Lincoln, U.S.A.}
\affiliation{Leibniz Institute for Polymer Research, Dresden, Germany}
\affiliation{Terahertz Materials Analysis Center, Department of Physics, Chemistry and Biology (IFM), Linköping University, SE 58183, Linköping, Sweden}

\date{}

\begin{abstract}

We determine the frequency dependence of the four independent Cartesian tensor elements of the dielectric function for monoclinic symmetry Y$_2$SiO$_5$ using generalized spectroscopic ellipsometry from 40-1200~cm$^{-1}$. Three different crystal cuts, each perpendicular to a principle axis, are investigated. We apply our recently described augmentation of lattice anharmonicity onto the eigendielectric displacement vector summation approach [A.~Mock~\textit{et al.}, Phys. Rev. B \textbf{95}, 165202 (2017)], and we present and demonstrate the application of an eigendielectric displacement loss vector summation approach with anharmonic broadening. We obtain excellent match between all measured and model calculated dielectric function tensor elements and all dielectric loss function tensor elements. We obtain 23 A$_{\mathrm{u}}$ and 22 B$_{\mathrm{u}}$ symmetry long wavelength active transverse and longitudinal optical mode parameters including their eigenvector orientation within the monoclinic lattice. We perform density functional theory calculations and obtain 23 A$_{\mathrm{u}}$ symmetry and 22 B$_{\mathrm{u}}$ transverse and longitudinal optical mode parameters and their orientation within the monoclincic lattice. We compare our results from ellipsometry and density functional theory and find excellent agreement. We also determine the static and above reststrahlen spectral range dielectric tensor values and find a recently derived generalization of the Lyddane-Sachs-Teller relation for polar phonons in monoclinic symmetry materials satisfied [M. Schubert, Phys. Rev. Lett. \textbf{117}, 215502 (2016)].

\end{abstract}
 \maketitle

\section {Introduction}
The optical properties of rare-earth ion doped single crystal materials have been the focus of substantial interest over the recent past. Their unique optical properties render  these materials highly suitable, for example, in optical applications as active laser mediums\cite{Li_1992, Sellin_1999, Thorpe_2011, Thiel_2014, Wang_2015, Cook_2015}, in optical signal processing\cite{Babbitt_2014, Saglamyurek_2014} and in quantum optics.\cite{Lvovsky_2009, Sangouard_2011, Bussieres_2013} Rare-earth Ce$^{3+}$ or Eu$^{3+}$ doped monoclinic yttrium orthosilicate (Y$_2$SiO$_5$) can be used as phosphorous material\cite{Sengthong_2016, Shin_2001, Meijerink_1991, Born_1985, Gomes_1969} or as scintillator material for detection of x-rays and $\gamma$-rays\cite{Nikl_2010}. Cr$^{4+}$ doped Y$_2$SiO$_5$ has been studied for use as saturable absorber in Q-switching laser devices.\cite{Kuo_1995, Deka_1992} Y$_2$SiO$_5$ has been investigated for use in quantum optical information technologies\cite{Thiel_2011, Turukhin_2001}. Pr$^{3+}$ doped Y$_2$SiO$_5$ was investigated for electromagnetically induced transparency\cite{Li_2016}.

Despite its wide use in visible spectral range optical applications a rather incomplete knowledge seems to exist about its accurate long-wavelength optical properties.  For example, a complete set of the transverse optical (TO) and longitudinal optical (LO) phonon mode frequencies, amplitudes, and eigendielectric displacement vectors has not been determined, neither by theory nor by experiment. Infrared (IR) spectra measurements and a tentative phonon band assignment was performed by Lazarev~\textit{et al.} (Ref.~\onlinecite{Lazarev_1963}). Raman investigations have been performed by Voron'ko~\textit{et al.} (Ref.~\onlinecite{Voronko_2011}) and by Zheng~\textit{et al.} (Ref.~\onlinecite{Zheng_2007}) Fourier transform IR (FTIR) spectroscopy analysis with incomplete TO mode assignment was performed recently by H\"{o}fer~\textit{et al.} (Ref. \onlinecite{Hofer_2016}). The LO mode parameters remain obscure thus far. To our best knowledge, no phonon mode calculations were performed for this material despite its intriguing visible spectral range optical properties.

\begin{figure}[hbt]
\centering
\includegraphics[width=0.5\textwidth]{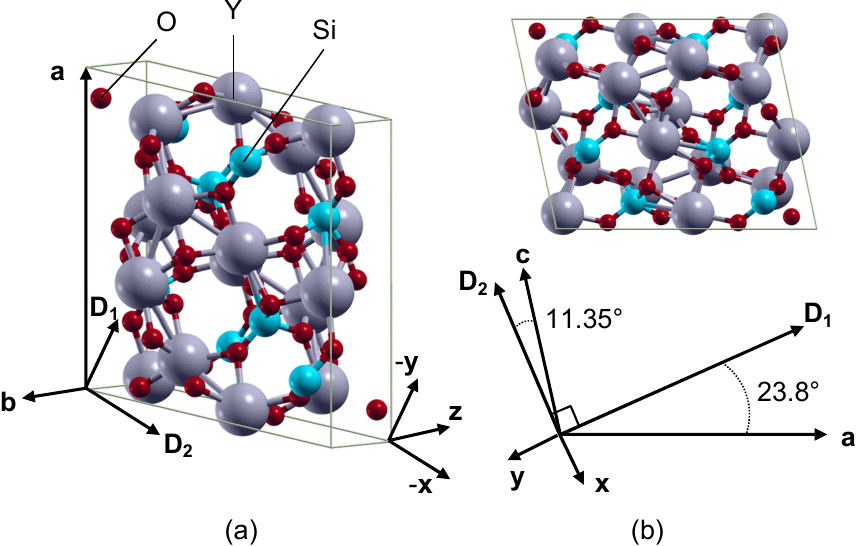}
\caption{\label{unitcell}(a) Unit cell of Y$_2$SiO$_5$, crystallographic axes (\textbf{a}, \textbf{b}, \textbf{c}), principal directions of the biaxial optical indicatrix (D$_1$, D$_2$, D$_3$) as defined in Ref.~\onlinecite{Li_1992} (D$_3$ is collinear with $\mathbf{b}$) and sample Cartesian coordinate system ($x$, $y$, $z$) used in this work. (b) View onto the $\mathbf{a}$ - $\mathbf{c}$ plane along axis $\mathbf{b}$, which points into the plane. The laboratory coordinate axes $x$ and $y$ are aligned with principle directions -D$_2$ and -D$_1$, respectively.}
\end{figure}

In this work, we provide a spectroscopic investigation of the long-wavelength anisotropic properties of Y$_2$SiO$_5$ by generalized spectroscopic ellipsometry (GSE). GSE is a convenient, contactless, non-destructive technique, which utilizes polarization of light transmitted through or reflected off an arbitrarily anisotropic sample allowing for the determination of both the real and imaginary parts of all nine complex dielectric function tensor elements. Recently, GSE has been used to characterize monoclinic materials. Jellison~\textit{et al.} first reported on determination of the dielectric function of a monoclinic crystal cadmium tungstate (CdWO$_4$ or CWO) using GSE in the spectral range of 1.5 to 4~eV and reported the need for four independent dielectric function tensor elements when describing the full spectral response of the monoclinic samples. This requirement differed from all previously GSE investigated anisotropic materials with orthorhombic, hexagonal, and tetragonal crystal symmetries where a maximum of three independent tensor elements sufficed.\cite{Fujiwara} Jellison~\textit{et al.} also reported on the determination of the four real values of the dielectric function tensor of the monoclinic crystal lutetium oxyorthosilicate (Lu$_2$SiO$_5$ or LSO) using GSE in the spectral range of 200 to 850 ~nm.\cite{Jellison_2012}

We have recently reported that an eigendielectric displacement vector summation (EDVS) approach can be used as a physical model approach to explain and line-shape match experimentally determined dielectric function tensor elements of materials with monoclinic and triclinic symmetries.\cite{Schubert_2016,Schubert_2016_LST} For long-wavelength excitations, the EDVS approach is equivalent to the microscopic
Born-Huang description of polar lattice vibrations in the harmonic approximation.\cite{Born_1954} The EDVS approach goes beyond the Born-Huang description because it provides access to the LO mode properties including their eigendielectric displacement loss directions. We applied the EDVS approach to monoclinic $\beta$-Ga$_2$O$_3$\cite{Schubert_2016,Mock_2017Ga2O3} and CdWO$_4$\cite{Mock_2017} and determined the complete set of transverse and longitudinal long-wavelength excitations including their directions within the monoclinic lattices. It was further shown that the EDVS approach leads to a revised formulation of the Lyddane-Sachs-Teller relation, \cite{Lyddane41} derived originally for isotropic materials, for materials with monoclinic and triclinic crystal symmetries. In the generalized-LST relation, the ratio of the determinants of the anisotropic static and high-frequency dielectric permittivity tensors is related to the squares of the ratios of all LO and TO mode frequencies, respectively.\cite{Schubert_2016_LST}

We have described recently the need to augment anharmonic lattice broadening effects onto the EDVS approach for correct match of measured dielectric function spectra from crystals with monoclinic symmetry. The anharmonic broadening was proposed for orthorhombic and higher symmetry materials by Berreman and Unterwald\cite{Berreman68} and Lowndes\cite{Lowndes70} (BUL broadening). We successfully demonstrated the augmentation of the BUL broadening for CdWO$_4$.\cite{Mock_2017}

In this work, we demonstrate that the EDVS approach can be used to also describe the complete dielectric loss response tensor for monoclinic materials. Thereby, we describe the eigendielectric displacement loss vector summation (EDLVS) approach. The EDLVS approach permits direct determination of the LO mode frequencies, broadening, amplitude, and eigenpolarization direction parameters. This approach dispenses with the need of numerical root finding algorithm in order to derive the LO mode frequencies from the EDVS approach. Both approaches, EDVS and EDLVS, while mathematically form invariant and interchangeable, provide useful access to physical parameters of TO and LO modes directly from measured quantities. We augment the same anharmonic broadening (BUL broadening) onto the EDVLS approach, and demonstrate excellent match between experimental and model calculated data sets for crystals of monoclinic Y$_2$SiO$_5$. Thereby, we identify and determine the full set of long-wavelength active phonon modes for Y$_2$SiO$_5$. In this paper, we discuss the results of multiple approaches, simultaneously performing best-match calculation procedures using complex-valued spectra of the determinant and the inverse determinant of the dielectric tensor, the dielectric tensor element spectra and the inverse dielectric tensor element spectra.

% \textcolor{red}{RAFAL: It is a good place to add a few words with regards to the situation of DFT calculation for lattice vibration modes in orthosilicates. The silicon tetroxide anion, SiO$^{4−}$ (silicate) is found in most mineral-forming silicate as oxoanions, for example, lutetium-yttrium oxyorthosilicate or gadolinium oxyorthosilicate. XXX  Here in this work we use DFT calculations XXX  and obtain all long-wavelength active phonon modes in YSO.}

In parallel with experimental studies, Y$_2$SiO$_5$ has been studied computationally using density functional theory. These studies were motivated by potential applications of the material, for example, in barrier coatings (and hence focused on mechanical and thermal properties, and defects\cite{Liu_2013,Luo_2014}); and as a host matrix for doping with rare-earth elements.\cite{Ching_2003} To the best of our knowledge, there is no comprehensive DFT study of phonons in Y$_2$SiO$_5$ available so far. It is worth noting that Y$_2$SiO$_5$ is isostructural with a number of rare earth silicates, from Dy$_2$SiO$_5$ to Lu$_2$SiO$_5$.\cite{Felsche_1973} Thus, Y$_2$SiO$_5$ can be a convenient model system for a range of other materials.

\section{Theory}
\subsection{Symmetry}

Monoclinic Y$_2$SiO$_5$ belongs to the space group 15 (centered monoclinic). International Tables for Crystallography\cite{ITA92} list 18 alternative choices of the unit cell for this space group and several of them has been used for Y$_2$SiO$_5$ in the literature. The crystallographic standard for monoclinic cells require choosing a cell with the shortest two translations in the net perpendicular to the symmetry direction $b$, with $c<a$, $\beta$ non-acute, and appropriate centering.\cite{kennard67,mighell02} In the case of Y$_2$SiO$_5$ these requirements are met by choosing the $I2/c$ cell, which we will consistently use throughout this paper. The structural parameters of the unit cell are discussed in the next section.

\subsection{Density Functional Theory}

\begin{table}
\caption{\label{tab:lattice} Comparison between the experimental and theoretical lattice constants (in \AA; monoclinic angle $\beta$ in $^{\circ}$).}
\begin{ruledtabular}
\begin{tabular}{lccccccc}
&Exp.$^\textrm{a}$&Exp.$^\textrm{b}$&Exp.$^\textrm{c}$&Exp.$^\textrm{d}$&Calc.$^\textrm{e}$&Calc.$^\textrm{f}$&Calc.$^\textrm{g}$\\
\hline
$a$&12.38&12.64&12.490&12.469&12.402&12.33&12.847\\
$b$&6.689&6.82&6.721&6.710&6.6149&6.594&6.807\\
$c$&10.34&10.52&10.410&10.388&10.237&10.23&10.722\\
$\beta$&102.53&102.50&102.65&102.68&101.98&102.2&107.15\\
\end{tabular}
\end{ruledtabular}
\begin{flushleft}
\footnotesize{$^\textrm{a}${Ref.~\onlinecite{harris65}.}}\\
\footnotesize{$^\textrm{b}${Ref.~\onlinecite{michel67}.}}\\
\footnotesize{$^\textrm{c}${Ref.~\onlinecite{maksimov68}.}}\\
\footnotesize{$^\textrm{d}${Ref.~\onlinecite{leonyuk99}, Cr doped.}}\\
\footnotesize{$^\textrm{e}${This work, LDA-PZ.}}\\
\footnotesize{$^\textrm{f}${Refs. \onlinecite{Liu_2013} and \onlinecite{Luo_2014}, LDA.}}\\
\footnotesize{$^\textrm{g}${Ref. \onlinecite{Ching_2003}, LDA-OLCAO.}}\\
\end{flushleft}
\end{table}

\begin{table}
\caption{\label{tab:structural} Calculated equilibrium structural parameters of Y$_2$SiO$_5$ determined in this work in comparison with selected literature values. Atomic positions are given in fractional coordinates of \textbf{a}, \textbf{b}, and \textbf{c} respectively. For the sake of consistency literature data from different sources has been converted to the same equivalent $I2/c$ cell and atomic positions and provided at the same level of accuracy.}
%\begin{center}
\begin{tabular}{lccc}
\hline
\hline
\noalign{\smallskip}
\multicolumn{4}{c}{Exp. (Ref.~\onlinecite{michel67})}\\
\hline
Y1&~0.463&~0.241&~0.432\\
Y2&~0.143&-0.380&~0.308\\
Si&-0.316&~0.414&~0.380\\
O1&~0.126&~0.287&~0.292\\
O2&~0.407&~0.492&~0.063\\
O3&~0.204&~0.372&~0.029\\
O4&~0.188&~0.094&-0.216\\
O5&~0.019&~0.415&-0.375\\
\noalign{\smallskip}\noalign{\smallskip}\noalign{\smallskip}
\multicolumn{4}{c}{Exp. (Ref.~\onlinecite{maksimov68})}\\
\hline
Y1&~0.463&~0.243&~0.429\\
Y2&~0.141&-0.378&~0.306\\
Si&-0.319&~0.407&~0.373\\
O1&~0.118&~0.287&~0.300\\
O2&~0.411&~0.498&~0.054\\
O3&~0.202&~0.343&~0.032\\
O4&~0.203&~0.071&-0.237\\
O5&~0.015&~0.398&-0.382\\
\noalign{\smallskip}\noalign{\smallskip}\noalign{\smallskip}
\multicolumn{4}{c}{Calc. (this work)}\\
\hline
Y1&~0.464&~0.244&~0.426\\
Y2&~0.139&-0.368&~0.307\\
Si&-0.318&~0.408&~0.371\\
O1&~0.117&~0.293&~0.300\\
O2&~0.413&~0.508&~0.058\\
O3&~0.202&~0.353&~0.029\\
O4&~0.201&~0.063&-0.246\\
O5&~0.019&~0.403&-0.381\\
\hline
\hline
\end{tabular}
%\end{center}
\end{table}

Theoretical calculations of long-wavelength active $\Gamma$-point phonon frequencies were performed by plane wave DFT using Quantum ESPRESSO (QE)\cite{[{Quantum ESPRESSO is available from http://www.qu\-an\-tum-es\-pres\-so.org. See also: }]GiannozziJPCM2009QE} We used the exchange correlation functional of Perdew and Zunger (PZ)\cite{PerdewPRB1981}. We employ optimized norm-conserving Vanderbilt (ONCV) scalar-relativistic pseudopotentials,\cite{Hamann2013} which we generated for the PZ functional using the code ONCVPSP\cite{ONCVPSP} with the optimized parameters of the SG15 distribution of pseudopotentials.\cite{SG15} The initial parameters of the unit cell and atomic positions were taken from Ref. \onlinecite{michel67}. The calculations were performed in a primitive cell \textbf{p$_1$} = \textbf{a}, \textbf{p$_2$} = \textbf{b}, \textbf{p$_3$} =(\textbf{a+b+c})/2 appropriate for the body-centered $I2/c$ cell. The conversions between equivalent cells and the preparation of the primitive cell were performed with the help of VESTA\cite{vesta} and CIF2CELL.\cite{cif2cell} The initial structure was first relaxed to force levels less than 10$^{-4}$ Ry Bohr$^{-1}$. A regular shifted $4\times4\times4$ Monkhorst-Pack grid was used for sampling of the Brillouin zone.\cite{MonkhorstPRBGRID} A convergence threshold of $1\times10^{-12}$  Ry was used to reach self consistency with a large electronic wavefunction cutoff of 100 Ry. The comparison of resulting optimized cell parameters with the existing literature data are listed in Tables \ref{tab:lattice} (unit cell parameters) and  \ref{tab:structural} (atomic positions). The relaxed cell was used for subsequent phonon calculations, which are described in Section \ref{dft_phonons}.

\begin{figure*}[hbt]
\centering
\includegraphics[width=.95\linewidth]{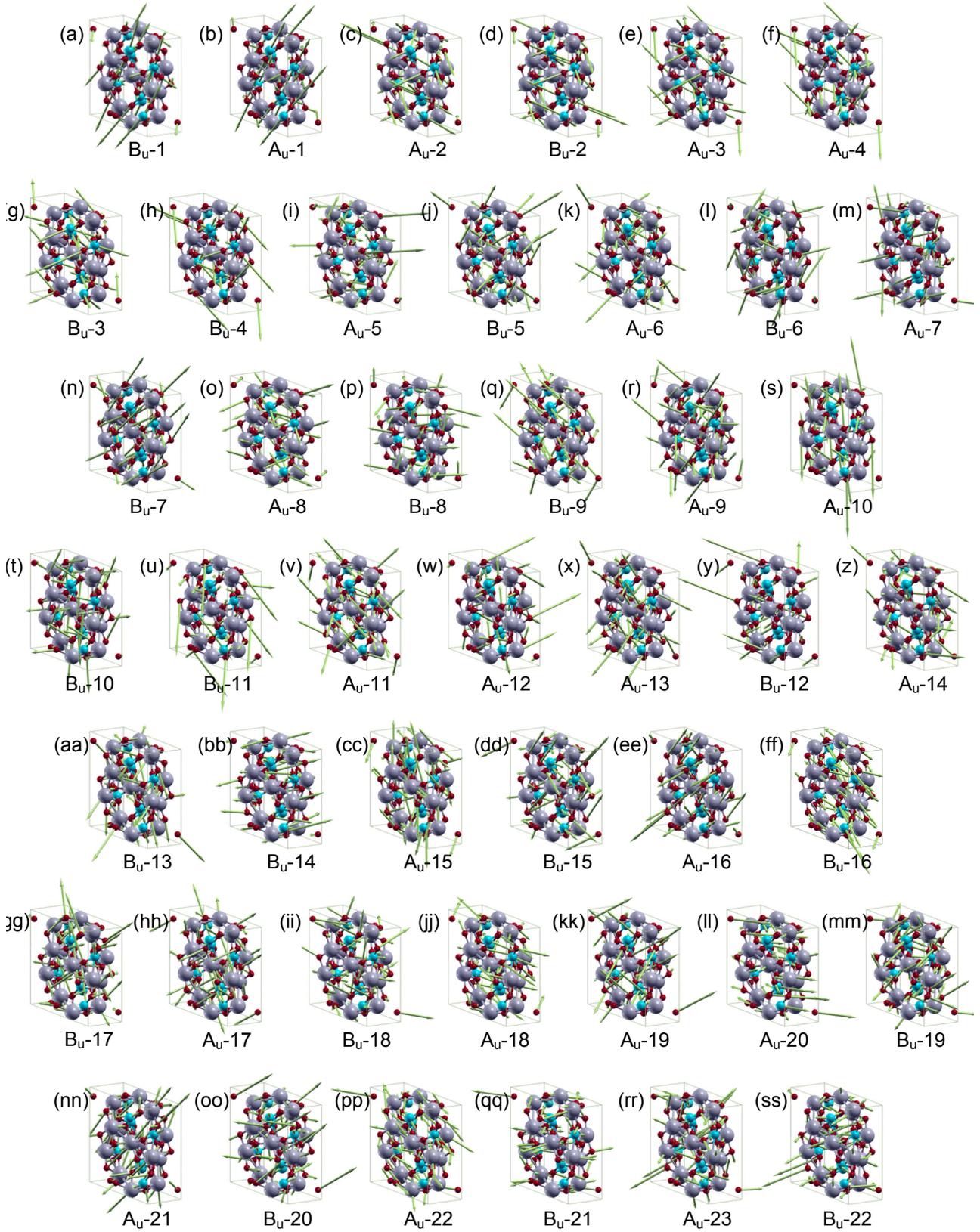}
\caption{\label{fig:phonons} Renderings of TO phonon modes in Y$_2$SiO$_5$ with $\mathrm{A_u}$ and $\mathrm{B_u}$ symmetry as labeled for each mode and presented in frequency order. The phonon mode frequency parameters were calculated using Quantum Espresso and are presented in Tabs. \ref{tab:BuDFT} and \ref{tab:AuDFT}. }
\end{figure*}

\subsection{TO and LO mode frequencies and vectors}

Two characteristic sets of eigenmodes can be defined from the frequency dependent dielectric function tensor, $\varepsilon(\omega)$, and dielectric loss function tensor, $\varepsilon^{-1}(\omega)$. These belong to the TO and LO modes. TO modes occur at frequencies in which dielectric resonance occurs for electric fields along $\mathbf{\hat{e}}_{l}$ with eigendielectric displacement unit vectors then defined as $\mathbf{\hat{e}}_{l} = \mathbf{\hat{e}}_{\mathrm{TO},l}$. Similarly, LO modes occur when the dielectric loss approaches infinity for electric fields along $\mathbf{\hat{e}}_{l}$ with eigendielectric displacement unit vectors then defined as $\mathbf{\hat{e}}_{l} = \mathbf{\hat{e}}_{\mathrm{LO},l}$. This can be written as:
\begin{subequations}\label{eq:TOLOvectors}
\begin{align}
|\det\{ \varepsilon(\omega=\omega_{\stext{TO},l})\}| &\rightarrow \infty, \\
|\det\{ \varepsilon^{-1}(\omega=\omega_{\stext{LO},l})\}| &\rightarrow \infty, \\
\varepsilon^{-1}(\omega=\omega_{\stext{TO},l})\mathbf{\hat{e}}_{\stext{TO},l} &=0,\\
\varepsilon(\omega=\omega_{\stext{LO},l})\mathbf{\hat{e}}_{\stext{LO},l} &=0,
\end{align}
\end{subequations}
where $l$ is an index for multiple frequencies in the sets.\cite{Schubert_2016_LST}

\subsubsection{The eigendielectric displacement approach}

\paragraph{$\varepsilon(\omega)$:} The EDVS approach can be used to best-match model calculate the dielectric function tensor of materials with monoclinic symmetry.\cite{Schubert_2016,Schubert_2016_LST,Mock_2017} The dielectric function tensor $\varepsilon$ is obtained from a sum of all contributions from individual dielectric resonances with displacement parallel to $\mathbf{\hat{e}}_{\mathrm{TO},l}$, added to a high-frequency scalar tensor $\varepsilon_{\infty}$. The latter accounts for all eigendielectric contributions from much shorter wavelengths,
\begin{equation}\label{eq:epssum}
\varepsilon=\varepsilon_\infty+\sum^{N}_{l=1}\varrho_{\mathrm{TO},l}(\mathbf{\hat{e}}_{\mathrm{TO},l}\otimes\mathbf{\hat{e}}_{\mathrm{TO},l}),
\end{equation}
where $\otimes$ is the dyadic product and $\varrho_{\mathrm{TO},l}$ are wavelength dependent functions that describe the responses of the $l=1,...,N$ long wavelength active TO displacement modes. In this approach, parameters in functions $\varrho_{\mathrm{TO},l}$ and directions $\mathbf{\hat{e}}_{\mathrm{TO},l}$ are directly accessible.

\paragraph{$\varepsilon^{-1}(\omega)$:} The EDLVS approach can be used to best-match model calculate the inverse dielectric function tensor of materials with monoclinic symmetry. The inverse dielectric function tensor $\varepsilon^{-1}$ is obtained from a sum of all contributions from individual dielectric loss resonances with displacement parallel to $\mathbf{\hat{e}}_{\mathrm{LO},l}$, added to a high-frequency scalar tensor $\varepsilon^{-1}_{\infty}$. The latter accounts for all eigendielectric loss contributions from much shorter wavelengths,
\begin{equation}\label{eq:epsinversesum}
\varepsilon^{-1}=\varepsilon^{-1}_\infty+\sum^{N}_{l=1}\varrho_{\mathrm{LO},l}(\mathbf{\hat{e}}_{\mathrm{LO},l}\otimes\mathbf{\hat{e}}_{\mathrm{LO},l}),
\end{equation}
where $\otimes$ is the dyadic product and $\varrho_{\mathrm{TO},l}$ are wavelength dependent functions that describe the responses of the $l=1,...,N$ long wavelength active LO displacement loss modes. In this approach, parameters in functions $\varrho^{-1}_{\mathrm{LO},l}$ and directions $\mathbf{\hat{e}}_{\mathrm{LO},l}$ are directly accessible. Without providing further proof, we state that the number of modes $N$ in Eqs.~\ref{eq:epssum} and ~\ref{eq:epsinversesum} must always equal.

\subsubsection{Model response functions}

We use anharmonic broadened Lorentzian oscillator functions to describe the TO and LO mode responses in Eqs.~\ref{eq:epssum} and Eqs.~\ref{eq:epsinversesum}, respectively.

\begin{equation}\label{eq:AHLO}
\varrho_{k,l} \left(\omega\right)=\frac{A_{k,l}^2-i\Gamma_{k,l}\omega}{\omega^2_{k,l}-\omega^2-i\omega\gamma_{k,l}}.
\end{equation}
\noindent Here, $A_{k,l}$, $\omega_{k,l}$, $\gamma_{k,l}$, and $\Gamma_{k,l}$ denote amplitude, resonance frequency, harmonic broadening, and anharmonic broadening parameter for TO ($k$=``TO'') or LO ($k$=``LO'') mode $l$, respectively, and $\omega$ is the frequency of the driving electromagnetic field.

\subsubsection{Coordinate-invariant generalized dielectric function with anharmonic broadening}

A factorized form of the dielectric function for long-wavelength active phonon modes was described by Berreman and Unterwald\cite{Berreman68} and by Lowndes\cite{Lowndes70} which allowed for determination of TO and LO mode frequencies in materials with multiple phonon modes. However, the Berreman-Unterwald-Lowndes (BUL) form was described under the assumption that all phonon modes which contributed to a dielectric function or inverse dielectric function under consideration must be polarized in the same crystal direction. Recently, a generalized coordinate-invarient approach was described by Schubert, Ref.~\onlinecite{Schubert_2016_LST} which discussed how the determinant of the dielectric function tensor could be utilized regardless of crystal symmetry:
\begin{equation}\label{eq:general-eps}
det\{\varepsilon(\omega)\}=det\{\varepsilon_\infty\}\prod_{l=1}^{N}\frac{\omega^2_{\stext{LO},l}-\omega^2}{\omega^2_{\stext{TO},l}-\omega^2}.
\end{equation}
This approach has recently been used by us for the analysis of monoclinic $\beta$-Ga$_2$O$_3$ and CdWO$_4$.\cite{Schubert_2016,Mock_2017} Note that the coordinate-invariant generalized dielectric function can reveal negative imaginary parts within distinct frequency intervals when the so-called ``TO-LO'' rule is broken in materials with monoclinic symmetry. We will discuss such occurrences in Section \ref{sec:TOLOrule}.

Eq. \ref{eq:AHLO} can be shown to directly transform into a BUL factorized form of the dielectric function equivalent to the four parameter semiquantum (FPSQ) model suggested by Gervais and Periou. The FPSQ model identifies $\gamma_{\mathrm{LO},l}$ to account for lifetime broadenings of LO modes different from those of associated TO modes, $\gamma_{\mathrm{TO},l}$.\cite{Gervais74} This four parameter model has been used for accurate description of the effects of anharmonic phonon mode coupling in anisotropic materials.\cite{Gervais74,SchoecheJAP2013TiO2,SchubertPRB61_2000,Schubert03g}

This inclusion of anharmonic broadening into the generalized coordinate-invarient generalized dielectric function, modifies Eq. \ref{eq:general-eps} into the form:
\begin{equation}\label{eq:general-eps-broaded}
det\{\varepsilon(\omega)\}=det\{\varepsilon_\infty\}\prod_{l=1}^{N}\frac{\omega^2_{\stext{LO},l}-\omega^2-i\omega\gamma_{\stext{LO},l}}{\omega^2_{\stext{TO},l}-\omega^2-i\omega\gamma_{\stext{TO},l}}.
\end{equation}

\subsubsection{Coordinate-invariant generalized dielectric loss function with anharmonic broadening}

A function analogous to Eq.~\ref{eq:general-eps-broaded} can be obtained for the dielectric loss response, and which has the following form
\begin{equation}\label{eq:general-epsinv-broaded}
det\{\varepsilon^{-1}(\omega)\}=det\{\varepsilon^{-1}_\infty\}\prod_{l=1}^{N}\frac{\omega^2_{\stext{TO},l}-\omega^2-i\omega\gamma_{\stext{TO},l}}{\omega^2_{\stext{LO},l}-\omega^2-i\omega\gamma_{\stext{LO},l}}.
\end{equation}

\subsubsection{Coordinate system for Y$_2$SiO$_5$}

For monoclinic Y$_2$SiO$_5$ we utilize the orthorhombic system D$_1$~$\times$~D$_2$~$\times$~b-axis (Fig. \ref{unitcell}) where D$_1$ and D$_2$ are parallel to the \textbf{a-c} plane and with our laboratory $x$ aligned with -D$_2$, $y$ aligned with -D$_1$, and $z$ aligned with axis $\mathbf{b}$.

\subsubsection{Dielectric function tensor model for Y$_2$SiO$_5$}

23 TO modes with $\mathrm{A_u}$ symmetry are polarized along axis $\mathbf{b}$. 22 TO modes with $\mathrm{B_u}$ symmetry are polarized within the \textbf{a-c} plane.
The dielectric tensor elements for Y$_2$SiO$_5$ are then rendered as:
\begin{subequations}\label{eq:epsmonoall}
\begin{align}
\varepsilon_{\mathrm{xx}} &= \varepsilon_{\infty,\mathrm{xx}}+\sum^{22}_{l=1}\varrho^{\mathrm{B_u}}_{\mathrm{TO},l}\cos^2\alpha_{\mathrm{TO},l},\\
\varepsilon_{\mathrm{xy}} &= \varepsilon_{\infty,\mathrm{xy}}+\sum^{22}_{l=1}\varrho^{\mathrm{B_u}}_{\mathrm{TO},l}\sin\alpha_{\mathrm{TO},l}\cos\alpha_{\mathrm{TO},l},\\
\varepsilon_{\mathrm{yy}} &= \varepsilon_{\infty,\mathrm{yy}}+\sum^{22}_{l=1}\varrho^{\mathrm{B_u}}_{\mathrm{TO},l}\sin^2\alpha_{\mathrm{TO},l},\\
\varepsilon_{\mathrm{zz}} &= \varepsilon_{\infty,\mathrm{zz}}+\sum^{23}_{l=1}\varrho^{\mathrm{A_u}}_{\mathrm{TO},l},\\
\varepsilon_{\mathrm{xz}} &= \varepsilon_{\mathrm{zx}} = \varepsilon_{\mathrm{zy}}= \varepsilon_{\mathrm{yz}}=0,
\end{align}
\end{subequations}

\noindent where angle $\alpha_{\mathrm{TO},l}$ denotes the orientation of the TO eigendielectric displacement vectors with $\mathrm{B_u}$ symmetry relative to laboratory coordinate \textbf{a} axis.

\subsubsection{Dielectric loss function tensor model for Y$_2$SiO$_5$}

23 LO modes with $\mathrm{A_u}$ symmetry are polarized along the axis $\mathbf{b}$. 22 LO modes with $\mathrm{B_u}$ symmetry are polarized within the \textbf{a-c} plane. The dielectric loss tensor elements for Y$_2$SiO$_5$ are then rendered as:
\begin{subequations}\label{eq:epsinvmonoall}
\begin{align}
\varepsilon^{-1}_{\mathrm{xx}} &= \varepsilon^{-1}_{\infty,\mathrm{xx}}+\sum^{22}_{l=1}\varrho^{\mathrm{B_u}}_{\mathrm{LO},l}\cos^2\alpha_{\mathrm{LO},l},\\
\varepsilon^{-1}_{\mathrm{xy}} &= \varepsilon^{-1}_{\infty,\mathrm{xy}}+\sum^{22}_{l=1}\varrho^{\mathrm{B_u}}_{\mathrm{LO},l}\sin\alpha_{\mathrm{LO},l}\cos\alpha_{\mathrm{LO},l},\\
\varepsilon^{-1}_{\mathrm{yy}} &= \varepsilon^{-1}_{\infty,\mathrm{yy}}+\sum^{22}_{l=1}\varrho^{\mathrm{B_u}}_{\mathrm{LO},l}\sin^2\alpha_{\mathrm{LO},l},\\
\varepsilon^{-1}_{\mathrm{zz}} &= \varepsilon^{-1}_{\infty,\mathrm{zz}}+\sum^{23}_{l=1}\varrho^{\mathrm{A_u}}_{\mathrm{LO},l},\\
\varepsilon^{-1}_{\mathrm{xz}} &= \varepsilon^{-1}_{\mathrm{zx}} = \varepsilon^{-1}_{\mathrm{zy}}= \varepsilon^{-1}_{\mathrm{yz}}=0,
\end{align}
\end{subequations}

\noindent where angle $\alpha_{\mathrm{LO},l}$ denotes the orientation of the TO eigendielectric displacement vectors with $\mathrm{B_u}$ symmetry relative to laboratory coordinate \textbf{a} axis.

\subsubsection{Complementary parameter analyses}
Eqs.~\ref{eq:epsmonoall} and~\ref{eq:epsinvmonoall}, augmented with response functions in Eq.~\ref{eq:AHLO}, are fully complementary, and one set of parameters ($\varepsilon_\infty$, $\mathrm{A}_{\mathrm{TO},l}$ $\omega_{\mathrm{TO},l}$, $\gamma_{\mathrm{TO},l}$, $\Gamma_{\mathrm{TO},l}$, $\mathbf{\hat{e}}_{\mathrm{TO},l}$) is in principle sufficient to determine the other set of parameters ($\varepsilon^{-1}_\infty$, $\mathrm{A}_{\mathrm{LO},l}$ $\omega_{\mathrm{LO},l}$, $\gamma_{\mathrm{LO},l}$, $\Gamma_{\mathrm{LO},l}$, $\mathbf{\hat{e}}_{\mathrm{LO},l}$). Analysis of experimental dielectric function data using Eq.~\ref{eq:epsmonoall} directly permits access to TO mode parameters, including their orientations. Analysis of experimental dielectric loss function data using Eqs.~\ref{eq:epsinvmonoall} directly permits access LO mode parameters, including their orientations. The immediate advantage of having wavelength-by-wavelength determined data for a dielectric function tensor available is to also have its inverse then available. The dielectric (loss) tensor elements reveal peak maxima in the imaginary parts that are directly associated with TO (LO) modes. Hence, one can read by ``eye inspection'' already from raw data where to anticipate TO and LO mode parameters.

Mayerh\"{o}fer~\textit{et al.} determined LO mode frequencies from modeling reflectance data of anisotropic materials parameterizing the dielectric function tensor using its inverse and the LO mode parameter set.\cite{Mayerhofer_2016} We have previously shown that LO mode frequencies in monoclinic materials can be determined by simply observing maxima in the inverse dielectric tensor.\cite{Schubert_2016} We have also shown that including this inverse tensor into the model analysis yields improved sensitivity to anharmonic broadening parameters.\cite{Mock_2017} In this work, we use both approaches and determine both sets of parameters, simultaneously analyzing dielectric function tensor and inverse dielectric function data.

\subsection{Generalized ellipsometry}

Generalized ellipsometry has been successfully used previously to investigate anisotropic materials including biaxial, uniaxial, and multilayered materials as well as metamaterials.\cite{SchubertPRB61_2000,SchoecheJAP2013TiO2,Schubert03g,DresselOE2008pentacene,Schubert03c,Ashkenov03,KasicPRB62_2000,KasicPRB65_2002,DarakchievaAPL84_2004,DarakchievaPRB2004AlN,DarakchievaPRB2005AlGaNSL,DarakchievaPRB2007StrainGaN,DarakchievaAPL2009InNe,DarakchievaAPL2009InN,DarakchievaAPL2010InN,DarakchievaPRB2014InNmix,DarakchievaJAP2014InNMg,HofmannTHzGLADChapter2013,Sekora_2016,mock_2016b,Briley_2017} Recently, it has been applied to monoclinic materials as well.\cite{JellisonPRB2011CdWO4,Schubert_2016_LST,Schubert_2016,Mock_2017,Mock_2017Ga2O3,Sturm_2015,Sturm_2016,Sturm_2017} Following the same approach used previously for $\beta$-Ga$_2$O$_3$\cite{Schubert_2016, Mock_2017Ga2O3} and CdWO$_4$\cite{Mock_2017}, data from multiple samples, multiple azimuths, multiple angle of incidences are investigated and analyzed simultaneously for Y$_2$SiO$_5$.

\subsubsection{Mueller matrix formalization}

In generalized ellipsometry, the Mueller matrix can be used to describe interaction of electromagnetic plane waves with anisotropic samples. Real-valued 4$\times$4 Mueller matrix elements are obtained which connect the Stokes vector components before and after interaction with the sample,
\begin{equation}
\left( {{\begin{array}{*{20}c}
 {S_{0} } \hfill \\ {S_{1} } \hfill \\  {S_{2} } \hfill \\  {S_{3} } \hfill \\
\end{array} }} \right)_{\mathrm{output}} =
\left( {{\begin{array}{*{20}c}
 {M_{11} } \hfill & {M_{12} } \hfill \ {M_{13} } \hfill & {M_{14} } \hfill \\
 {M_{21} } \hfill & {M_{22} } \hfill \ {M_{23} } \hfill & {M_{24} } \hfill \\
 {M_{31} } \hfill & {M_{32} } \hfill \ {M_{33} } \hfill & {M_{34} } \hfill \\
 {M_{41} } \hfill & {M_{42} } \hfill \ {M_{43} } \hfill & {M_{44} } \hfill \\
\end{array} }} \right)
\left( {{\begin{array}{*{20}c}
 {S_{0} } \hfill \\ {S_{1} } \hfill \\  {S_{2} } \hfill \\  {S_{3} } \hfill \\
\end{array} }} \right)_{\mathrm{input}}.
\end{equation}
with the Stokes vector components defined by $S_{0}=I_{p}+I_{s}$, $S_{1}=I_{p} - I_{s}$, $S_{2}=I_{45}-I_{ -45}$, $S_{3}=I_{\sigma + }-I_{\sigma - }$. Here, $I_{p}$, $I_{s}$, $ I_{45}$, $I_{-45}$, $I_{\sigma + }$, and $I_{\sigma - }$denote the intensities for the $p$-, $s$-, +45$^{\circ}$, -45$^{\circ}$, right handed, and left handed circularly polarized light components, respectively~\cite{Fujiwara_2007}.

\subsubsection{Wavelength-by-wavelength analysis}

In order to extract physical parameters, data must be analyzed through a best match model calculation procedure. We apply a half-infinite, two phase model to Y$_2$SiO$_5$ where two half-infinite mediums, ambient (air) and monoclinic Y$_2$SiO$_5$, are separated by the planar surface of the crystal.\cite{Schubert96,Schubert03a,SchubertIRSEBook_2004,Schubert04,Fujiwara_2007} In this approach, the Euler angles describing the orientation of the crystal axes and the elements of the monoclinic dielectric tensor are considered free parameters. The dielectric function tensor elements are expressed as wavelength dependent model functions, thereby allowing for determination of the tensor elements in the so-called wavelength-by-wavelength model analysis approach.

We establish two Cartesian coordinate systems such that our sample coordinate system may be related to the crystallographic axes or the so-called principle directions of the biaxial optical indicatrix of Y$_2$SiO$_5$.\cite{Traum_2014} The laboratory coordinate system is determined by the ellipsometer instrument and is defined by the sample holder and plane of incidence. The sample surface is defined as the $\hat{x}$ - $\hat{y}$ plane, and the sample normal defines the $\hat{z}$ axis which points into the sample. We assign the sample system ($x, y, z$) to coincide with the axes of the optical indicatrix (D$_1$, D$_2$, D$_3$), as defined in Ref.~\onlinecite{Li_1992}. Note that D$_3$ coincides with -$\mathbf{b}$ (Fig.~\ref{unitcell}), where we follow the notation given in Ref.~\onlinecite{Li_1992}. Due to the monoclinic symmetry the dielectric tensor, $\varepsilon$, for Y$_2$SiO$_5$, contains shear elements, and with the choice of coordinates above can now be expressed as
\begin{equation}\label{eq:monoclinicepsDC}
\boldsymbol{\varepsilon}= \left(
\begin{array}{ccc}
\varepsilon_{\mathrm{xx}} & \varepsilon_{\mathrm{xy}} & 0 \\
\varepsilon_{\mathrm{xy}} & \varepsilon_{\mathrm{yy}} & 0 \\
0          & 0 & \varepsilon_{\mathrm{zz}}

\end{array}\right).
\end{equation}

A Euler angle rotation can be applied to $\varepsilon$ in order to describe the crystallographic surface and azimuthal orientation of the sample. The sample azimuth, $\varphi$, defined by an in-plane rotation with respect to sample normal, describes the mathematical rotation of a model dielectric function tensor of calculated data when compared with measured data taken at different azimuthal orientations.

In a wavelength-by-wavelength approach, calculated Mueller matrix data is compared to experimentally measured Mueller matrix data. Wavelength dependent dielectric function tensor elements $\varepsilon_{\mathrm{xx}}$, $\varepsilon_{\mathrm{yy}}$, $\varepsilon_{\mathrm{xy}}$, and $\varepsilon_{\mathrm{zz}}$ are varied in order to minimized the mean square error ($\xi$) function.\cite{SchubertPRB61_2000,Schubert03h,SchubertIRSEBook_2004,SchubertADP15_2006,SchoecheJAP2013TiO2} Analysis of all samples, azimuthal orientations, and angles of incidence is performed simultaneously for all independent wavelengths yielding a single set of complex valued, wavelength dependent $\varepsilon_{\mathrm{xx}}$, $\varepsilon_{\mathrm{yy}}$, $\varepsilon_{\mathrm{xy}}$, and $\varepsilon_{\mathrm{zz}}$ (polyfit). During the polyfit, an independent set of Euler angle parameters for each sample is utilized to describe the orientation of the principle directions and crystallographic axes at the first azimuthal position acquired.

\subsubsection{Model analysis procedure}

% \textcolor{red}{Need to reword nicely... I'm struggling now with explaining this clearly}

% a-c plane: Analysis description...all parts performed simultaneously

In order to reduce correlation and improve sensitivity to model parameters, multiple data sets are fit simultaneously with multiple models. The model process is detailed below for the \textbf{a-c} plane with three parts. The model procedure is repeated independently for modes along the \textbf{b} axis

\paragraph{Model 1:} Eq.~\ref{eq:general-eps-broaded} and Eq.~\ref{eq:general-epsinv-broaded} are used to best-match model calculate the wavelength-by-wavelength determined determinants of $\varepsilon$ and $\varepsilon^{-1}$, respectively, finding parameters $\omega_{\mathrm{TO},l}$, $\gamma_{\mathrm{TO},l}$, $\omega_{\mathrm{LO},l}$, $\gamma_{\mathrm{LO},l}$, and $\varepsilon_\infty$. The best-match model calculated functions are represented by black solid lines in Fig.~\ref{fig:det}.

\paragraph{Model 2:} In addition to best-match model calculate determinants of $\varepsilon$ and $\varepsilon^{-1}$, the individual wavelength-by-wavelength determined dielectric tensor elements $\varepsilon_{\mathrm{xx}}$, $\varepsilon_{\mathrm{xy}}$, and $\varepsilon_{\mathrm{yy}}$ are best-match model calculated using Eqs.~\ref{eq:epsmonoall} (a), (b), and (c), respectively, and the anharmonic Lorentzian oscillator functions in Eq.~\ref{eq:AHLO} to determine the additional TO mode parameters A$_{\mathrm{TO},l}$, $\Gamma_{\mathrm{TO},l}$, and $\alpha_{\mathrm{TO},l}$. In addition, the numerically calculated inverse of the model calculated dielectric function tensor is matched with the wavelength-by-wavelength determined inverse of the dielectric function tensor. The best-match model calculated functions are represented by red solid lines in Figs. \ref{fig:tensor} and \ref{fig:inversetensor}.

\paragraph{Model 3} In addition to best-match model calculate determinants of $\varepsilon$ and $\varepsilon^{-1}$, the individual wavelength-by-wavelength determined inverse dielectric tensor elements $\varepsilon^{-1}_{\mathrm{xx}}$, $\varepsilon^{-1}_{\mathrm{xy}}$, and $\varepsilon^{-1}_{\mathrm{yy}}$ are best-match model calculated using Eqs.~\ref{eq:epsinvmonoall} (a), (b), and (c), respectively, and the anharmonic Lorentzian oscillator functions in Eq.~\ref{eq:AHLO} to determine the additional LO mode parameters A$_{\mathrm{LO},l}$, $\Gamma_{\mathrm{LO},l}$, and $\alpha_{\mathrm{LO},l}$. In addition, the numerically calculated inverse of the model calculated inverse dielectric function tensor is matched with the wavelength-by-wavelength determined dielectric function tensor. The best-match model calculated functions are represented by cyan solid lines in Figs.~\ref{fig:tensor} and~\ref{fig:inversetensor}.

\section{Experiment}

Three single crystal samples of Y$_2$SiO$_5$ purchased from Scientific Materials Corporation were investigated. The sample dimensions were 10~mm~$\times$~10~mm~$\times$~1~mm. Investigated crystal orientations were D$_1$ $\times$~D$_2$~$\times$~b-axis, b-axis~$\times$~D$_2$~$\times$~D$_1$, and D$_1$~$\times$~b-axis~$\times$~D$_2$, where primary axes D1 and D2 are defined in relation to the crystal axes as shown in Fig. \ref{unitcell} taken from Ref. \onlinecite{Li_1992} All model calculations were performed using WVASE32$^{TM}$ (J. A. Woollam Co., Inc.).

\section{Results and Discussion}

\subsection{DFT Phonon Calculations}\label{dft_phonons}

The phonon frequencies and transition dipole components were computed at the $\Gamma$-point of the Brillouin zone using density functional perturbation theory.~\cite{BaroniRMP2001DFTPhonons} The parameters of the TO modes were taken directly from the $\Gamma$-point calculations. The parameters of the LO modes were obtained by setting a small displacement from the $\Gamma$-point. For $\mathrm{A_u}$ symmetry modes this displacement was in the direction of the $\mathbf{b}$ axis. For the $\mathrm{B_u}$ modes, the entire $\mathbf{a} - \mathbf{c}$ plane was probed with a step of 10$^{\circ}$, and the parameters of the LO modes were taken at the direction, for which the angular dependence of the mode frequency for each $\mathrm{B_u}$ mode had its maximum value.

The results of the phonon mode calculations for all long wavelength active modes with $\mathrm{A_u}$ and $\mathrm{B_u}$ symmetry ($\omega_{\mathrm{TO},l}$, A$_{\mathrm{TO},l}$, $\alpha_{\mathrm{TO},l}$, $\omega_{\mathrm{LO},l}$, A$_{\mathrm{LO},l}$, $\alpha_{\mathrm{LO},l}$) are listed in Tabs.~\ref{tab:AuDFT} and \ref{tab:BuDFT}. Note that for modes with $\mathrm{A_u}$ symmetry, all eigenvectors are oriented along axis $\mathbf{b}$ and thus $\alpha_{\mathrm{TO,LO},l}$ are not defined. Values for $\alpha_{\mathrm{TO,LO},}$ for modes with $\mathrm{B_u}$ symmetry are counted relative to axis $\mathbf{a}$ within the $\mathbf{a} - \mathbf{c}$ plane. Renderings of atomic displacements for each mode were prepared using XCrysDen~\cite{XCrysDen} running under Silicon Graphics Irix 6.5, and are shown in Fig.~\ref{fig:phonons}.

\begin{table*}
\caption{\label{tab:BuDFT} DFT results for phonon modes with B$_{\mathrm{u}}$ symmetry in units of wavenumbers (cm$^{-1})$, Debye (D), Angstrom (\AA), angular degrees ($^{\circ}$) and atomic mass units (amu).}
\begin{ruledtabular}
\begin{tabular}{{l}{c}{c}{c}{c}{c}{c}}
Mode & $\omega_{\mathrm{TO},l}$ [cm$^{-1}$] & A$^2_{\mathrm{TO},l}$ [(D/\AA)$^2$/amu] & $\alpha_{\mathrm{TO},l}$ [$^\circ$] & $\omega_{\mathrm{LO},l}$ [cm$^{-1}$] & A$^2_{\mathrm{LO},l}$ [(D/\AA)$^2$/amu] & $\alpha_{\mathrm{LO},l}$ [$^\circ$]\\
\hline
1 & 966.11 & 84.81 & 29.47 & 1045.52 & 132.64 & 26.9 \\
2 & 891.27 & 32.54 & -44.46 & 951.88 & 118.87 & 120 \\
3 & 856.41 & 44.44 & -63.85 & 875.32 & 7.41 & 100 \\
4 & 853.56 & 1.90 & -74.88 & 853.71 & 0.072 & 50 \\
5 & 558.53 & 32.50 & 49.40 & 631.13 & 65.6 & 51 \\
6 & 527.01 & 10.12 & -53.59 & 558.38 & 35.24 & 140 \\
7 & 502.50 & 39.43 & 43.20 & 526.81 & 9.62 & 40 \\
8 & 493.58 & 3.30 & -75.69 & 500.62 & 21.59 & 135 \\
9 & 446.32 & 14.93 & -71.93 & 477.84 & 13.46 & 130 \\
10 & 406.86 & 4.07 & 61.75 & 425.91 & 16.41 & 170 \\
11 & 380.05 & 24.87 & -23.84 & 406.86 & 4.08 &  150\\
12 & 323.99 & 38.31 & -79.44 & 364.37 & 7.2 & 80 \\
13 & 309.05 & 25.65 & -4.52 & 323.98 & 40.77 & 195 \\
14 & 303.80 & 2.43 & 67.12 & 304.91 & 0.1758 & 80 \\
15 & 268.70 & 4.92 & 69.25 & 274.7 & 0.8884 & 80 \\
16 & 244.06 & 31.34 & -48.75 & 268.23 & 3.4168 & 160 \\
17 & 233.69 & 12.55 & 13.39 & 244.01 & 34.83 & 45 \\
18 & 219.59 & 8.19 & 30.63 & 224.05 & 0.1979 & 50 \\
19 & 166.79 & 1.15 & -72.86 & 168.38 & 0.1258 & 100 \\
20 & 150.81 & 2.24 & 79.28 & 153.94 & 0.2055 & 70 \\
21 & 108.39 & 0.85 & -34.66 & 110.01 & 0.0731 & 150 \\
22 & 43.66 & 0.98 & 71.93 & 49.11 & 0.1278 & 70 \\
\end{tabular}
\end{ruledtabular}
\end{table*}

\begin{table*}
\caption{\label{tab:AuDFT} Same as Tab.~\ref{tab:BuDFT} for A$_{\mathrm{u}}$ symmetry.}
\begin{ruledtabular}
\begin{tabular}{{l}{c}{c}{c}{c}}
Mode & $\omega_{\mathrm{TO},l}$ [cm$^{-1}$]& A$^2_{\mathrm{TO},l}$ [(D/\AA)$^2$/amu] & $\omega_{\mathrm{LO},l}$ [cm$^{-1}$] & A$^2_{\mathrm{LO},l}$ [(D/\AA)$^2$/amu] \\
\hline
1 & 956.65 & 1.80 & 977.36 & 2.11 \\
2 & 904.34 & 48.11 & 952.03 & 0.81 \\
3 & 872.12 & 25.36 & 881.09 & 0.39 \\
4 & 864.40 & 1.38 & 864.66 & 0.05 \\
5 & 589.15 & 12.84 & 615.54 & 1.14 \\
6 & 546.78 & 6.14 & 555.53 & 0.51 \\
7 & 526.02 & 0.06 & 526.11 & 0.06 \\
8 & 499.19 & 0.80 & 501.68 & 0.38 \\
9 & 426.55 & 13.91 & 473.15 & 1.08 \\
10 & 418.49 & 0.58 & 418.86 & 0.06 \\
11 & 379.78 & 0.73 & 400.33 & 0.53 \\
12 & 354.79 & 26.59 & 379.03 & 0.09 \\
13 & 339.33 & 14.90 & 344.73 & 0.11 \\
14 & 315.95 & 5.49 & 319.40 & 0.12 \\
15 & 274.18 & 1.90 & 279.51 & 0.25 \\
16 & 250.40 & 14.03 & 266.87 & 0.23 \\
17 & 229.70 & 2.77 & 233.56 & 0.12 \\
18 & 218.35 & 10.43 & 224.89 & 0.10 \\
19 & 208.79 & 0.02 & 208.81 & 0.01 \\
20 & 182.75 & 8.59 & 190.28 & 0.13 \\
21 & 154.73 & 1.14 & 155.86 & 0.05 \\
22 & 112.79 & 0.45 & 113.64 & 0.04 \\
23 & 104.07 & 0.49 & 104.91 & 0.04 \\
\end{tabular}
\end{ruledtabular}
\end{table*}

\subsection{Mueller matrix analysis}

\begin{figure*}[hbt]
\centering
\includegraphics[width=\linewidth]{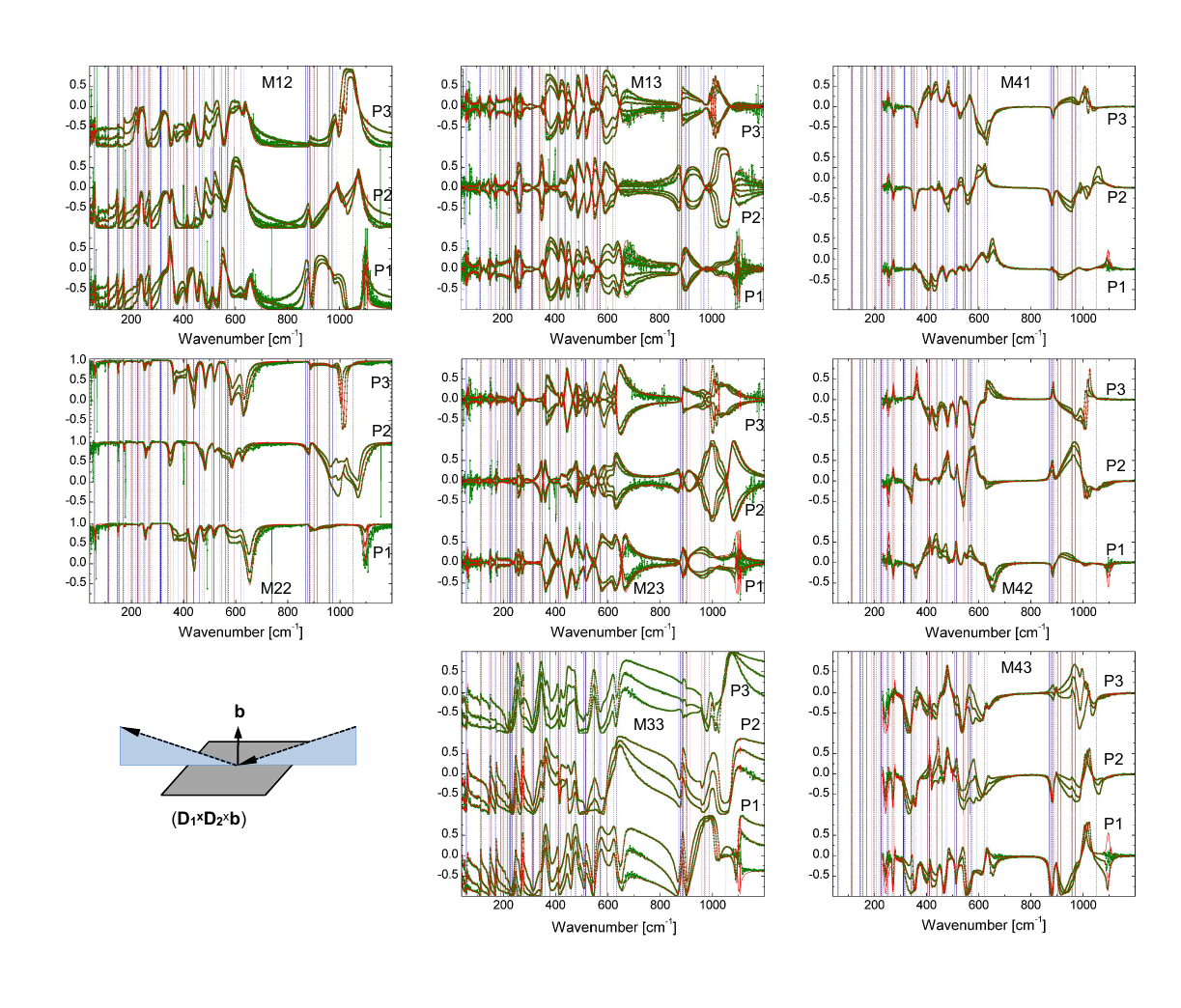}
\caption{\label{fig:MM008}Experimental (dotted, green lines) and best match model calculated (solid, red lines) Mueller matrix data obtained from a (D$_1\times$D$_2\times$\textbf{b}) surface at three representative sample azimuth orientations. (P1: $\varphi=3.6(3)^{\circ}$, P2: $\varphi=48.6(3)^{\circ}$, P3: $\varphi=93.6(3)^{\circ}$). Data were taken at three angles of incidence ($\Phi_a=50^{\circ}, 60^{\circ}, 70^{\circ}$). Equal Mueller matrix data, symmetric in their indices, are plotted within the same panels for convenience. Vertical lines indicate wavenumbers of TO (solid lines) and LO (dotted lines) modes with $\mathrm{B_u}$ symmetry (blue) and $\mathrm{A_u}$ symmetry (brown). Note that the fourth column elements are plotted as the fourth column for convenience. Fourth row elements are only available from the IR instrument limited to approximately 230~cm$^{-1}$ and are plotted in the symmetric tensor panel locations for convenience. Note that all elements are normalized to M$_{11}$. The remaining Euler angle parameters are $\theta=0.2(4)$ and $\psi=-0.1(5)$ consistent with the crystallographic orientation of the (D$_1\times$D$_2\times$\textbf{b}) surface. The inset depicts schematically the sample surface, the plane of incidence, and the orientation of axis $\mathbf{b}$ in P1.}
\end{figure*}

\begin{figure*}[hbt]
\centering
\includegraphics[width=\linewidth]{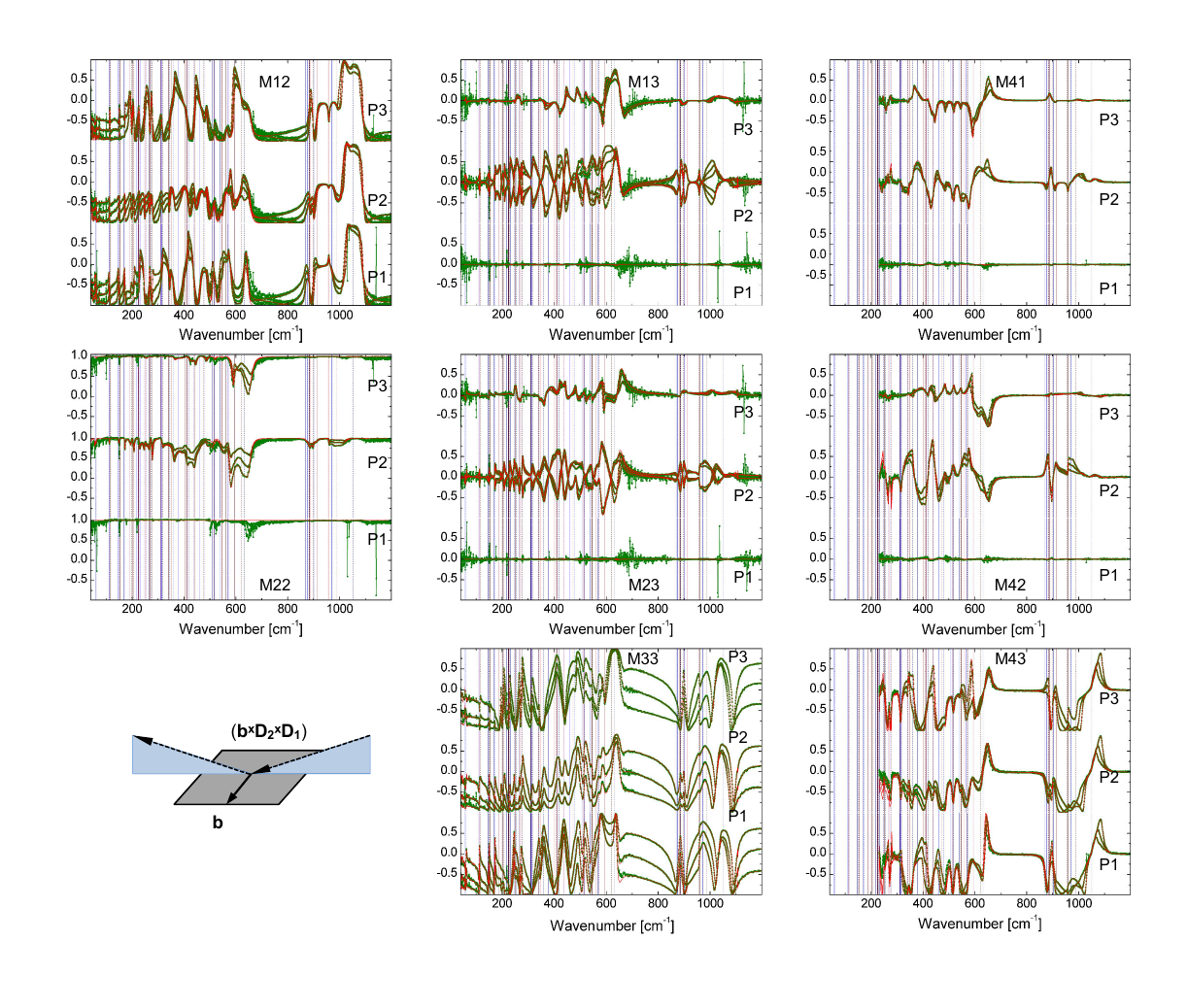}
\caption{\label{fig:MM006}Same as Fig.~\ref{fig:MM008} for the (\textbf{b}$\times$D$_1\times$D$_2$) sample at azimuth orientation P1: $\varphi=-0.2(1)^{\circ}$, P2: $\varphi=44.7(9)^{\circ}$, P3: $\varphi=89.79^{\circ}$. $\theta=89.(9)$ and $\psi=-1.5(7)$, consistent with the crystallographic orientation of the (\textbf{b}$\times$D$_1\times$D$_2$) surface. Note that in position P1, axis $\mathbf{b}$ which is parallel to the sample surface in this crystal cut, is aligned almost perpendicular to the plane of incidence. Hence, the monoclinic plane with $\mathbf{a}$ and $\mathbf{c}$ is nearly parallel to the plane of incidence, and as a result almost no conversion of $p$ to $s$ polarized light occurs and vice versa. As a result, the off diagonal block elements of the Mueller matrix are near zero. The inset depicts schematically the sample surface, the plane of incidence, and the orientation of axis $\mathbf{b}$, shown approximately for position P1.}
\end{figure*}

Figs. \ref{fig:MM008} and \ref{fig:MM006} show representative experimental and best match model calculated Mueller matrix data for two of the three surfaces investigated in this work, namely the (D$_1\times$D$_2\times$\textbf{b}) and (\textbf{b}$\times$D$_1\times$D$_2$) surfaces. Insets in Figs.~\ref{fig:MM008} and~\ref{fig:MM006} show schematically the $\mathbf{b}$ axis with respect to the sample surface, and the plane of incidence is also indicated. Individual panels are shown for each Mueller matrix element and are arranged according to Mueller matrix indices. Within each panel, data from 3 different azimuthal positions (P1, P2 and P3), each 45$^\circ$ rotated clockwise, each with 3 angles of incidence (50$^\circ$, 60$^\circ$, and 70$^\circ$) are presented. Data from additional positions measured are not shown for brevity.

It is observed by experiment as well as by model calculations that all Mueller matrix elements are symmetric, i.e., $\mathrm{M}_{ij}=\mathrm{M}_{ji}$, therefore, symmetric elements i.e., from upper and lower diagonal parts of the Mueller matrix, are plotted within the same panels. Therefore, only panels from the upper part of a $4\times4$ matrix arrangement is presented, and because all data obtained are normalized to element $\mathrm{M}_{11}$, the first column does not appear in this arrangement. Element $\mathrm{M}_{44}$ cannot be obtained in our current instrument configuration due to the lack of a second compensator and is therefore not presented. Data are shown for wavenumbers from 40 cm$^{-1}$ to 1200 cm$^{-1}$, except for row $\mathrm{M}_{4j}=\mathrm{M}_{j4}$ which only contains data from approximately 250 cm$^{-1}$ to 1200 cm$^{-1}$ because the fourth row is unavailable with our FIR instrumentation. Note that the fourth row data is plotted in the fourth column of Figs. \ref{fig:MM008} and \ref{fig:MM006} for convenience. All other panels show data obtained within the FIR range (40 cm$^{-1}$ to 500 cm$^{-1}$) using our FIR instrumentation and data obtained within the IR range (500 cm$^{-1}$ to 1200 cm$^{-1}$) using our IR instrumentation.

Strong anisotropy is noted in Y$_2$SiO$_5$ by the non-zero contributions in off-block diagonal elements (M$_{13}$, M$_{23}$, M$_{14}$, and M$_{24}$) and a strong dependence on azimuthal orientation is also apparent by inspection of the Mueller matrix data. Another important observation from the Mueller matrix data is that at P1 for the (\textbf{b}$\times$D$_1\times$D$_2$) surface in Fig. \ref{fig:MM006}, where the \textbf{b} axis is parallel to the sample surface and perpendicular to the plane of incidence, the off-block diagonal elements are very nearly zero. This is because the monoclinic plane is parallel to the plane of incidence in this orientation and therefore there is no mode conversion of $s$-polarized light to $p$-polarized light and vice versa.

All data sets, while unique, share similar characteristic features at specific wavenumbers which are indicated by vertical lines. Below, we identify these vertical lines as frequencies of all TO and LO phonon mode with $\mathrm{A_u}$ and $\mathrm{B_u}$ symmetries. Analysis of all data sets were performed simultaneously, where for each wavelength up to 792 independent data points from the multiple samples, azimuthal positions and angles of incidence are included in the polyfit. Only 17 independent parameters are included as variables in this so-called wavelength-by-wavelength analysis, including the 8 real and imaginary parts of the dielectric tensor elements ($\varepsilon_{\mathrm{xx}}$, $\varepsilon_{\mathrm{yy}}$, $\varepsilon_{\mathrm{xy}}$, and $\varepsilon_{\mathrm{zz}}$) as well as 3 sets of wavelength independent Euler angles to describe the sample surface and orientation. The resulting Mueller matrix rendered from this polyfit analysis is shown in Figs. \ref{fig:MM008} and \ref{fig:MM006} as red solid lines, and resulting real and imaginary parts of the dielectric tensor elements are given in Fig. \ref{fig:tensor} as green dotted lines. We find excellent agreement between our measured experimental and model calculated Mueller matrix data, and the Euler angles determined by this analysis are consistent with the anticipated sample surfaces and crystallographic orientations.

\subsection{Dielectric tensor analysis}

\begin{figure*}[hbt]
\centering
\includegraphics[width=\textwidth]{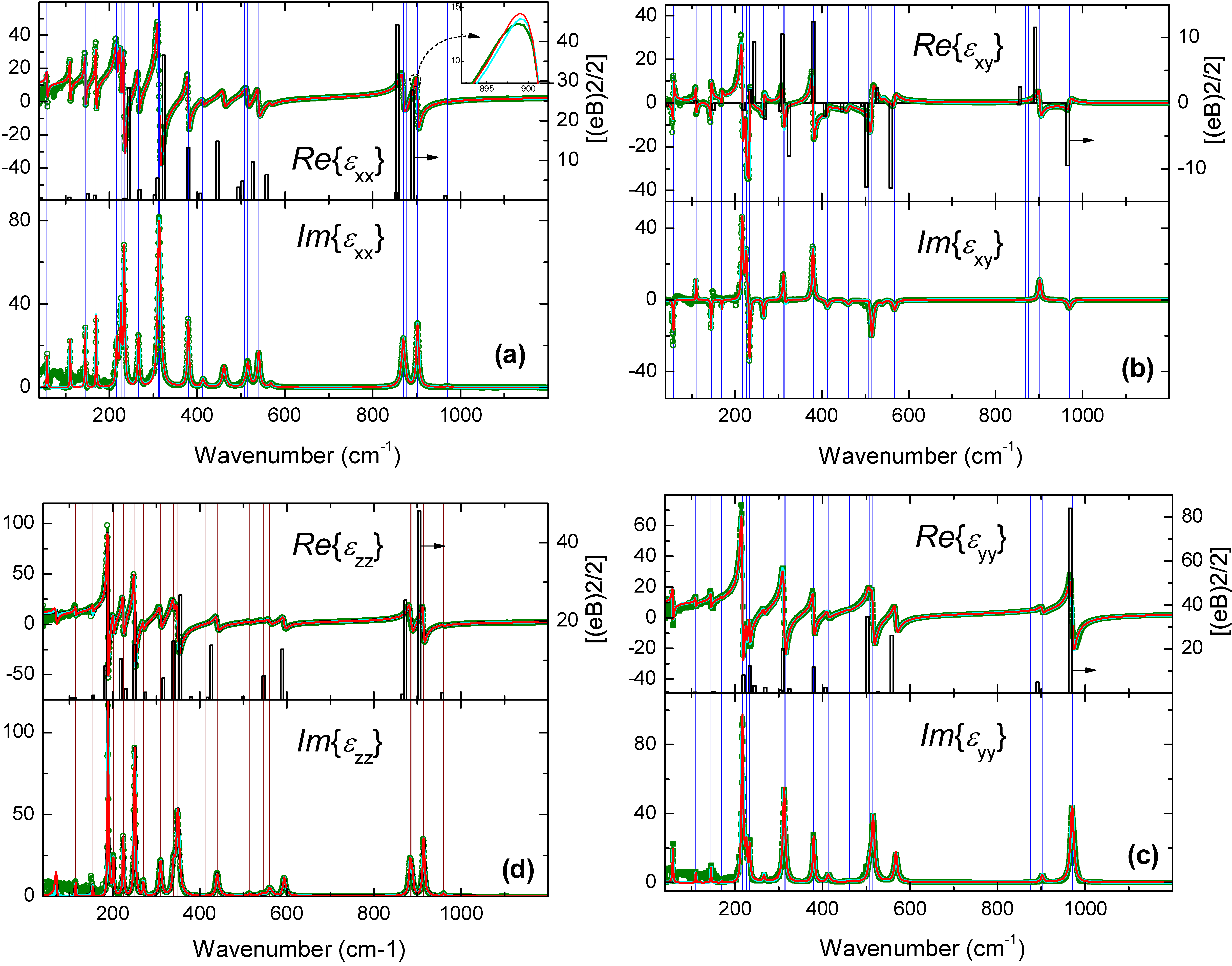}
\caption{\label{fig:tensor}Dielectric function tensor element $\varepsilon_{\mathrm{xx}}$ (a), $\varepsilon_{\mathrm{xy}}$ (b), $\varepsilon_{\mathrm{yy}}$ (c), and $\varepsilon_{\mathrm{zz}}$ (d). Green dotted lines indicate results from wavelength-by-wavelength best match model regression analysis matching the  experimental Mueller matrix data shown in Figs.~\ref{fig:MM008} and~\ref{fig:MM006}. Solid red lines are obtained from best match model lineshape analysis using Eqs.~\ref{eq:epsmonoall} with Eq.~\ref{eq:AHLO} fit to the dielectric tensor elements. Solid cyan lines are obtained from best match model lineshape analysis using a second set of Eqs.~\ref{eq:epsmonoall} with Eq.~\ref{eq:AHLO} fit to the inverse dielectric tensor elements.  Vertical lines in panel group [(a), (b), (c)], and in panel (d) indicate TO frequencies with $\mathrm{B_u}$ (blue) and $\mathrm{A_u}$ (brown) symmetry, respectively. Vertical bars indicate DFT calculated long-wavelength transition dipole moments in atomic units projected onto axis $x$, $y$, and $z$ as well as onto the shear plane $xy$.}
\end{figure*}

\begin{figure*}[hbt]
\centering
\includegraphics[width=.9\textwidth]{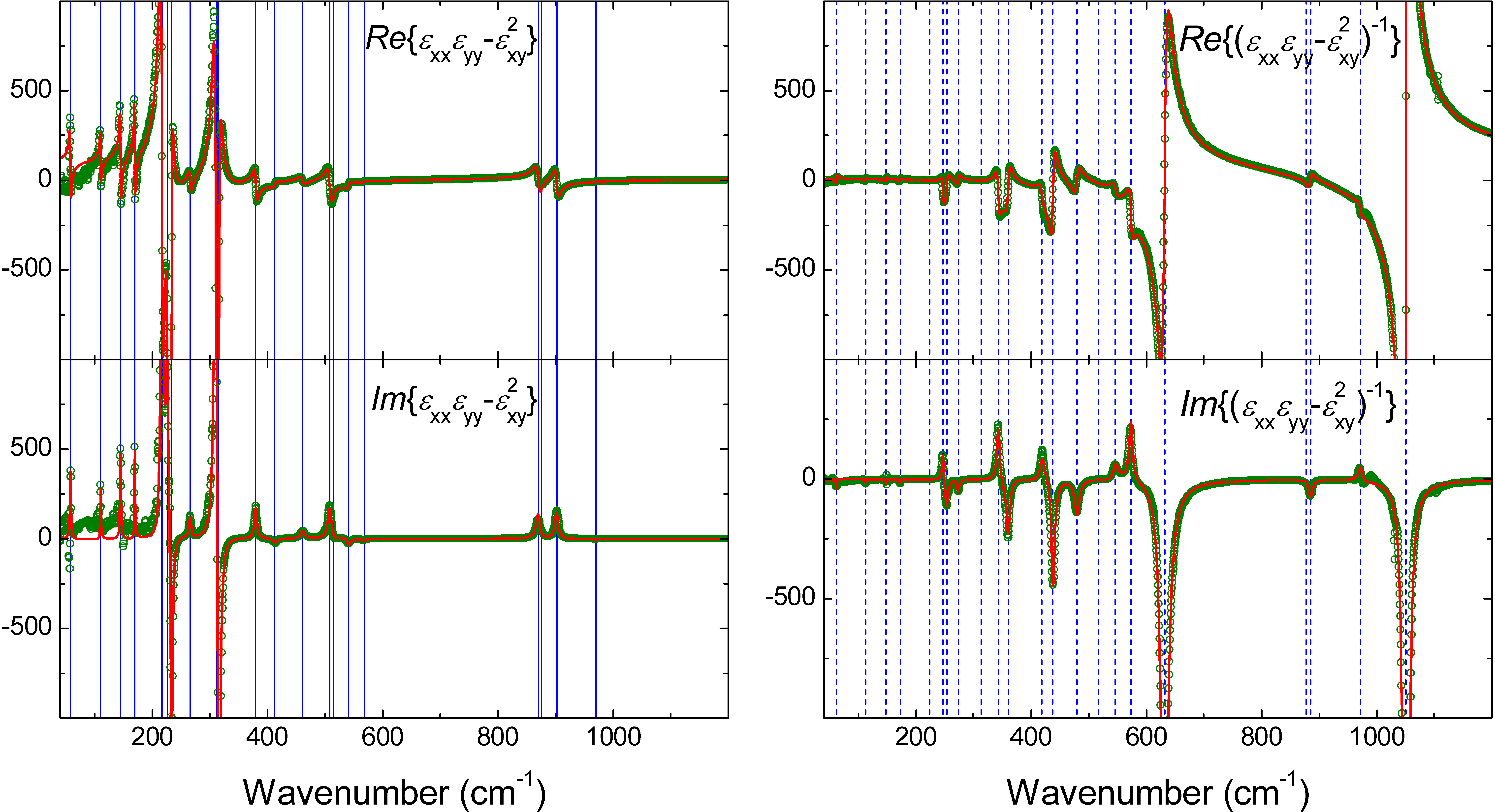}
\caption{\label{fig:det}(a) Real and imaginary parts of the coordinate invariant generalized dielectric function, $\varepsilon_{\mathrm{xx}}\varepsilon_{\mathrm{yy}}$-$\varepsilon^2_{\mathrm{xy}}$, along with its inverse (b), ($\varepsilon_{\mathrm{xx}}\varepsilon_{\mathrm{yy}}$-$\varepsilon^2_{\mathrm{xy}}$)$^{-1}$. Best match model calculated data (red, solid lines) calculated from the BUL form agrees excellently with data determined from a wavelength-by-wavelength analysis. TO and LO mode parameters are determined independent of their individual polarization and amplitudes. Frequencies of TO modes are indicated with solid blue lines and frequencies of LO modes are indicated by dotted blue lines.}
\end{figure*}

\begin{figure*}[hbt]
\centering
\includegraphics[width=.9\linewidth]{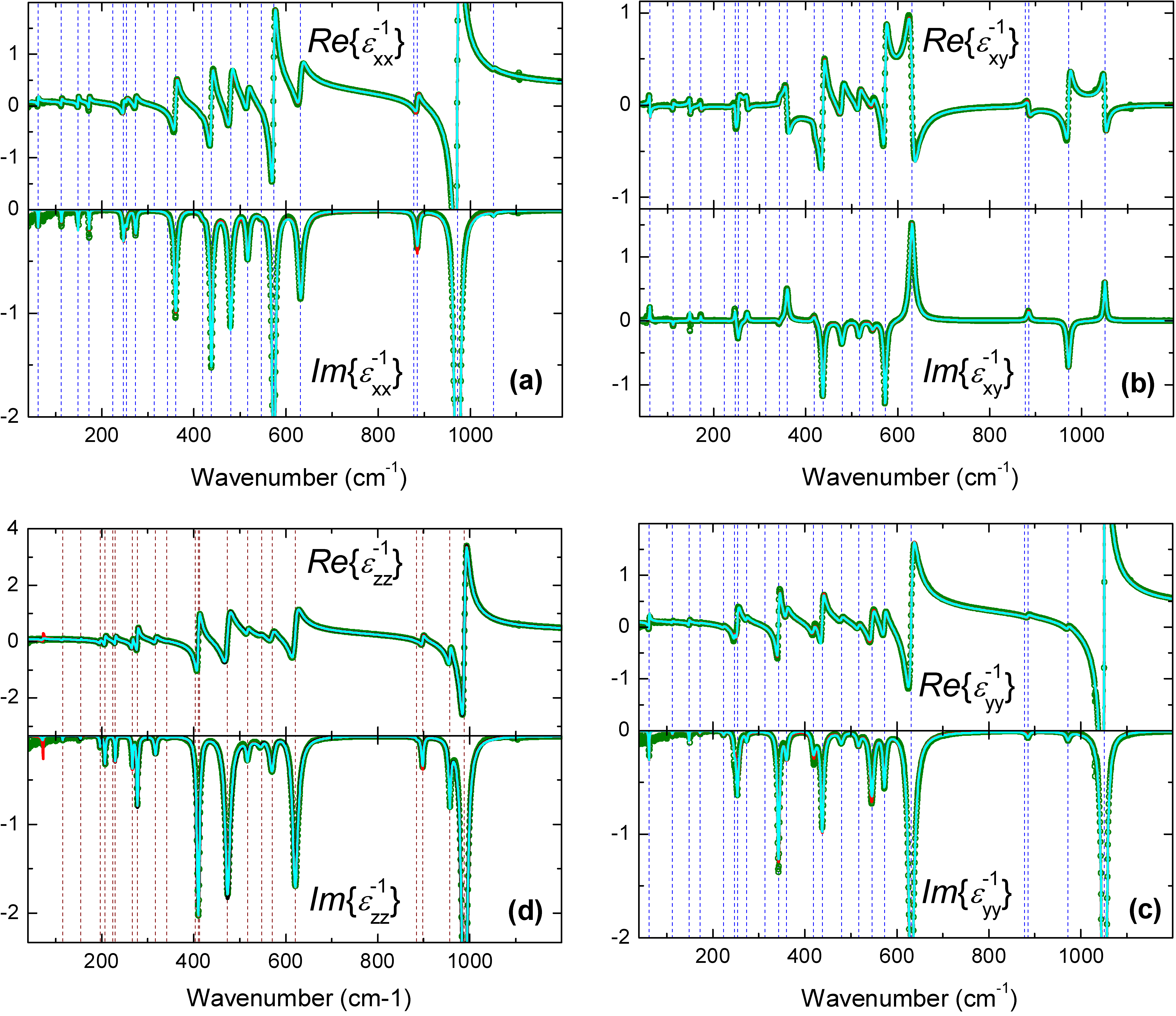}
\caption{\label{fig:inversetensor}Same as for Fig. \ref{fig:tensor} but for the inverse dielectric tensor. Vertical lines in panel group [(a), (b), (c)], and in panel (d) indicate LO frequencies with $\mathrm{B_u}$ (blue) and $\mathrm{A_u}$ (brown) symmetry, respectively.}
\end{figure*}

Real and imaginary parts of the dielectric tensor elements determined by the wavelength-by-wavelength polyfit are given in Fig.~\ref{fig:tensor} as green dotted lines for $\varepsilon_{\mathrm{xx}}$, $\varepsilon_{\mathrm{xy}}$, $\varepsilon_{\mathrm{yy}}$, and $\varepsilon_{\mathrm{zz}}$. One can then translate these into the inverse dielectric tensor shown as green dotted lines in Fig.~\ref{fig:inversetensor} for $\varepsilon^{-1}_{\mathrm{xx}}$, $\varepsilon^{-1}_{\mathrm{xy}}$, $\varepsilon^{-1}_{\mathrm{yy}}$, and $\varepsilon^{-1}_{\mathrm{zz}}$, and into the determinant  ($\varepsilon_{\mathrm{xx}}\varepsilon_{\mathrm{yy}}-\varepsilon_{\mathrm{xy}}^2$) and inverse determinant (($\varepsilon_{\mathrm{xx}}\varepsilon_{\mathrm{yy}}-\varepsilon_{\mathrm{xy}}^2)^{-1}$) as shown by green dotted lines in Fig.~\ref{fig:det}. From these, preliminary observations can be made for phonon mode properties. As we have previously reported, TO mode frequencies can be found from maxima in imaginary parts of the dielectric function tensor elements as well as the determinant\cite{Schubert_2016_LST} as shown in Figs. \ref{fig:tensor} and \ref{fig:det} indicated by solid vertical lines. Likewise, LO mode frequencies can be determined from maxima of the imaginary parts of the inverse dielectric function tensor and the inverse of the determinant\cite{Mock_2017} as seen in Figs. \ref{fig:det} and \ref{fig:inversetensor} indicated by dotted vertical lines. We note that panels (a), (b) and (c) in Figs. \ref{fig:tensor} and \ref{fig:inversetensor} share common frequencies at which maxima occur from which we identify 22 TO modes and a corresponding 22 LO modes with $\mathrm{B_u}$ symmetry, respectively. We also note that the imaginary part of $\varepsilon_{\mathrm{xy}}$ and $\varepsilon^{-1}_{\mathrm{xy}}$ can be positive as well as negative extrema at $\mathrm{B_u}$ TO and LO mode frequencies, respectively, which is due to the respective eigendielectric displacement unit vector orientation relative to axis $\mathbf{a}$. From Eq.~\ref{eq:epsmonoall}(b), it is seen that the imaginary part of $\varepsilon_{\mathrm{xy}}$ is negative when $\mathbf{\alpha}_{\stext{TO},l}$ is within $\{0 \dots -\pi \}$ and positive when $\mathbf{\alpha}_{\stext{TO},l}$ is within $\{0 \dots \pi \}$. Therefore, for example, we observe from experiment that $\mathrm{B_u}$ TO modes labeled 2, 3, 4, 6, 9, 11, 12, 13, 15, 17, 18, 19, and 21 are all oriented with negative angle towards axis $\mathbf{a}$.

% \begin{figure}[h]
% \centering
% \includegraphics[width=\linewidth]{exx_v1.png}
% \caption{\label{fig:exx}}
% \end{figure}

% \begin{figure}[h]
% \centering
% \includegraphics[width=\linewidth]{eyy_v1.png}
% \caption{\label{fig:eyy}}
% \end{figure}

% \begin{figure}[h]
% \centering
% \includegraphics[width=\linewidth]{exy_v1.png}
% \caption{\label{fig:exy}}
% \end{figure}

% \begin{figure}[h]
% \centering
% \includegraphics[width=\linewidth]{ezz_v1.png}
% \caption{\label{fig:ezz}}
% \end{figure}

% \begin{figure}[h]
% \centering
% \includegraphics[width=\linewidth]{invexx_v1.png}
% \caption{\label{fig:invexx}}
% \end{figure}

% \begin{figure}[h]
% \centering
% \includegraphics[width=\linewidth]{inveyy_v1.png}
% \caption{\label{fig:inveyy}}
% \end{figure}

% \begin{figure}[h]
% \centering
% \includegraphics[width=\linewidth]{invexy_v1.png}
% \caption{\label{fig:invexy}}
% \end{figure}

% \begin{figure}[h]
% \centering
% \includegraphics[width=\linewidth]{invezz_v1.png}
% \caption{\label{fig:invezz}}
% \end{figure}

\subsection{Phonon mode analysis}

\subsubsection{Modes with $\mathrm{B_u}$ symmetry in the \textbf{a-c} plane}

\paragraph{TO mode parameter determination} Solid red lines in Figs. \ref{fig:tensor} and \ref{fig:inversetensor} indicate the resulting best match model calculations obtained from Eq. \ref{eq:epsmonoall} using a set of anharmonically broadened Lorentzian oscillators. We find excellent agreement between our wavelength-by-wavelength and model calculated $\varepsilon$ and $\varepsilon^{-1}$. All best match TO model parameters are summarized in Tab. \ref{Table:ExpBu} including amplitude (A$_{\mathrm{TO},l}$), frequency ($\omega_{\mathrm{TO},l}$), broadening ($\gamma_{\mathrm{TO},l}$), anharmonic broadening ($\Gamma_{\mathrm{TO},l}$), and eigenvector direction ($\alpha_{\mathrm{TO},l}$) for all TO modes ($l=1...22$) with $\mathrm{B_u}$ symmetry. Frequencies of the TO modes are indicated by solid vertical blue lines in Figs. \ref{fig:MM008}, \ref{fig:MM006}, \ref{fig:tensor}, and \ref{fig:det} which align with the features observed in the data and the extrema seen in the imaginary part of the dielectric tensor.

TO mode parameters determined by H\"{o}fer \textit{et al.} (Ref. \onlinecite{Hofer_2016}) are included in Tab. \ref{Table:ExpBu} for comparison. While we do expect 22 modes with $\mathrm{B_u}$ symmetry from calculations, and they do identify 22 features, it can be seen that several features determined by H\"{o}fer \textit{et al.} do not correspond with modes determined by our analysis, specifically features identified at 974.3, 539.6, 461.5, and 231.2~cm$^{-1}$. In addition, several modes determined in our work are not identified by H\"{o}fer \textit{et al.}, specifically, modes 4, 14, 21 and 22.

\begin{table*}\centering
\caption{\label{Table:ExpBu}Phonon mode parameters with $\mathrm{B_u}$ symmetry obtained from best match model analysis of tensor element spectra $\varepsilon$ and $\varepsilon^{-1}$, using anharmonic broadened Lorentz oscillator functions in Eq.~\ref{eq:AHLO} as well as by utilization of the generalized coordinate invarient form of the dielectric function\cite{Schubert_2016_LST}. The last digit, which is determined within the 90\% confidence interval, is indicated with brackets for each parameter. Parameters for frequency and TO eigenvector orientation from Ref. \onlinecite{Hofer_2016} are given for comparison with orientation directions shifted by approximately $77^\circ$.}
\begin{ruledtabular}
\begin{tabular}{{l}|{c}{c}{c}{c}{c}{c}{c}{c}{c}{c}|{c}{c}}
 & \scriptsize{$\omega_{\mathrm{TO}}$(cm$^{-1}$)}&\scriptsize{$\omega_{\mathrm{LO}}$(cm$^{-1}$)}& \scriptsize{$\gamma_{\mathrm{TO}}$(cm$^{-1}$)}&\scriptsize{$\gamma_{\mathrm{LO}}$(cm$^{-1}$)}&\scriptsize{A$_{\mathrm{TO}}$(cm$^{-1}$)}&\scriptsize{$\Gamma_{\mathrm{TO}}$(cm$^{-1}$)}&\scriptsize{$\alpha_{\mathrm{TO}}$($^\circ$)}&\scriptsize{A$_{\mathrm{LO}}$(cm$^{-1}$)}&\scriptsize{$\Gamma_{\mathrm{LO}}$(cm$^{-1}$)}&\scriptsize{$\alpha_{\mathrm{LO}}$($^\circ$)}&$\bar{v}_j$\cite{Hofer_2016} &$\Phi_j$\cite{Hofer_2016} \\
\hline
- & &&&&&&&&& & 974.3 & 106.1 \\
1 & 970.5(9) & 1050.8(0) & 9.0(6) & 5.8(3) & 627.(2) & 14.(2) & 28.8(7) & 251.(1) & 0.3(6) & 27.2(6) & 970.7 & 107.1 \\
2 & 902.1(3) & 972.1(1) & 6.5(1) & 8.1(1) & 449.(8) & -8.(0) & -47.(0) & 240.(5) & -0.9(2) & 119.5(5) & 902.5 & 26.5 \\
3$^a$ & 876.(0) & 877.(0) & 5.(0) & 5.(2) & 99.(7) & 0.(0) & -68.(7) & 7.(4) & 0.(0) & -3(1) & 871.0 & 5.8 \\
4 & 869.9(0) & 885.(2) & 8.8(1) & 7.(9) & 42(0) & 13.(7) & 112.(7) & 52.(4) & 0.2(4) & -87.(8) &- & - \\
5 & 567.8(8) & 631.4(9) & 8.(0) & 13.1(8) & 294.(0) & 4.(0) & 40.(5) & 173.(9) & -2.8(0) & 52.4(1) & 567.9 & 115.5\\
6 & 540.2(6) & 573.1(2) & 6.(7) & 7.1(6) & 246.(9) & 5.(2) & 108.5(7) & 125.(5) & -0.8(0) & 135.0(3)& 540.2 & 107.4 \\
- & &&&&&&&&& &539.6&9.8\\
7 & 515.3(5) & 546.(1) & 8.0(4) & 10.(5) & 44(8) & -1(2) & 52.(2) & 60.(3) & 0.0(8) & -166.(7) &514.0&126.5\\
8 & 507.7(6) & 516.6(6) & 8.(1) & 8.(3) & 19(9) & -0.(8) & 3.(4) & 50.(0) & -0.3(5) & -39.(1)&507.9&54.7 \\
9 & 460.9(3) & 479.6(7) & 8.(7) & 9.(6) & 205.(4) & -3.(5) & 100.(0) & 75.(6) & 0.5(3) & 130.(7) &462.3&12.3\\
- & &&&&&&&&&&461.5&145.9  \\
10 & 412.(7) & 437.8(4) & 7.(9) & 7.2(9) & 16(1) & -(9) & 6(5) & 88.(2) & -0.7(6) & -27.(6)&413.1&144.4 \\
11 & 379.7(1) & 418.(6) & 5.0(9) & 10.(3) & 334.(6) & 2.(2) & -2(4) & 32.(5) & -0.6(0) & -132.(6)&379.5&53.5 \\
12 & 314.5(8) & 360.3(2) & 7.(0) & 7.7(0) & 3(2)5 & -3(0) & -7(8) & 58.(3) & 0.2(3) & -93.(1)&313.4&14.0 \\
13 & 312.1(0) & 343.0(6) & 6.2(8) & 5.(9) & 37(0) & (8) & -17.(3) & 49.9(6) & 0.0(5) & 21.(2)&312.6&102.6 \\
14 & 312.(2) & 313.(2) & 7.(3) & 6.(9) & 2(6)0 & 1(8) & -107.(8) & 2.7(8) & 0.0(1) & 23(2)&-&- \\
15 & 266.0(1) & 273.2(7) & 4.1(7) & 4.(7) & 176.(1) & -(4) & -87.(6) & 19.8(1) & -0.0(6) & -97.(7)&265.8&-9.8 \\
16 & 233.5(9) & 253.6(0) & 3.6(0) & 6.8(2) & 267.(8) & -1(4) & -95.(3) & 36.1(2) & -0.26(4) & -1.(2)&231.3&141.5 \\
- & &&&&&&&&&&231.2&4.6 \\
17 & 226.0(9) & 246.9(3) & 4.6(8) & 4.3(1) & 238.(7) & 2(9) & -3(2) & 21.0(2) & -0.02(4) & -101.(5) &225.4&63.2\\
18 & 216.4(7) & 223.9(2) & 4.5(5) & 5.6(7) & 339.(4) & -5(2) & -2.1(9) & 6.7(3) & 0.03(7) & -151.(3)&217.1&77.9 \\
19 & 169.6(2) & 172.1(8) & 1.5(8) & 2.1(0) & 98.(9) & -4.(9) & -75.(9) & 8.9(5) & -0.04(9) & -85.(5) & 170.9&0.6\\
20 & 144.9(8) & 148.4(2) & 2.0(8) & 2.0(1) & 106.(0) & -0.(3) & -96.(1) & 9.4(3) & 0.01(5) & -101.(1)&145.0&-18.3 \\
21 & 110.4(2) & 112.2(2) & 1.5(6) & 1.8(6) & 70.(4) & -2.(7) & -40.(7) & 5.7(6) & -0.02(4) & -38.(5)&-&- \\
22 & 57.8(9) & 61.7(1) & 2.1(2) & 1.6(9) & 65.(3) & 3.(9) & -115.(9) & 7.0(4) & 0.04(8) & -118.(5) &-&-\\
		\end{tabular}
\end{ruledtabular}
\begin{flushleft}
\footnotesize{$^\textrm{a}${Mode parameters fit in a local region, held constant in full spectral fit procedure.}}\\
\end{flushleft}
\end{table*}

\paragraph{TO eigendielectric displacement vectors}

\begin{figure}[hbt]
\centering
	\includegraphics[width=\linewidth]{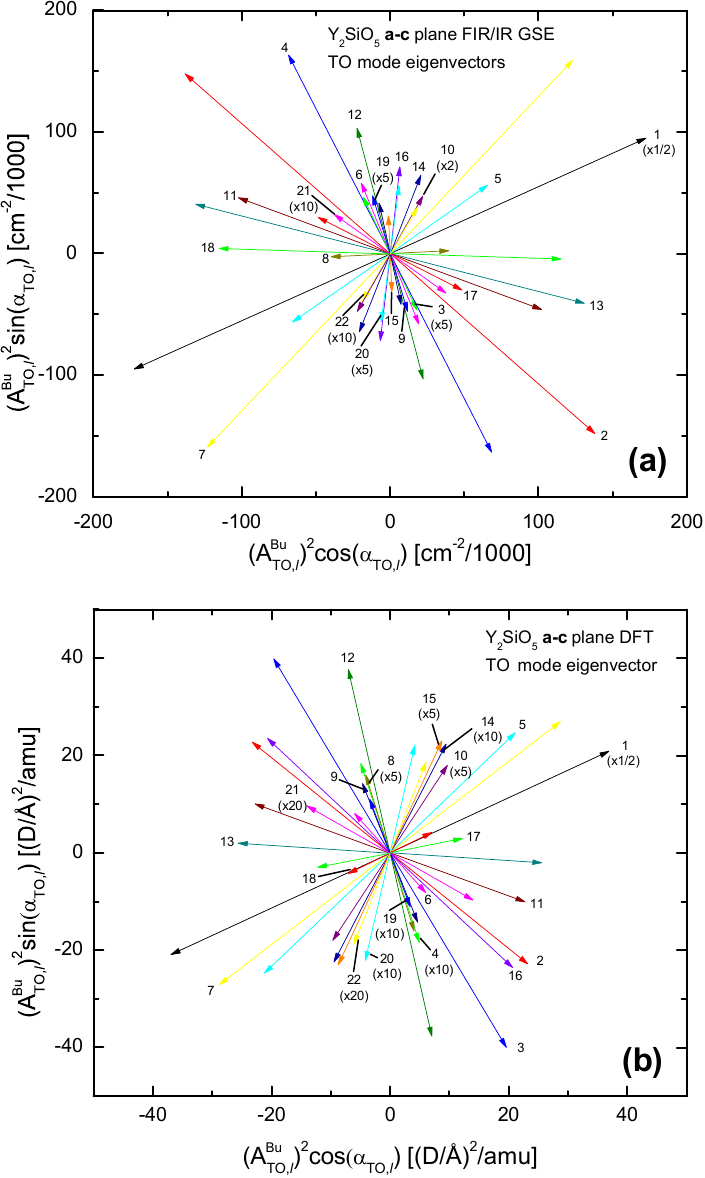}
\caption{\label{fig:TOvectors} (a) Schematic representation of the eigendielectric displacement vectors with GSE analysis determined amplitude $A_{\mathrm{TO},k}^{\mathrm{Bu}}$ and orientation angle $\alpha_{\mathrm{TO},k}$ (with respect to the \textbf{a} axis) of TO modes with $\mathrm{B_u}$ symmetry within the \textbf{a-c} plane. (b) DFT calculated long-wavelength transition dipoles (intensities) of TO modes with $\mathrm{B_u}$ symmetry. }
\end{figure}

Fig. \ref{fig:TOvectors} displays a vector representation of the amplitude and polarization direction parameters (A$^{\mathrm{Bu}}_{\mathrm{TO},l}$ and $\alpha_{\mathrm{TO},l}$) within the \textbf{a-c} plane. Results from the IR/FIR GSE model dielectric function, panel (a), are compared with long-wavelength transition dipole moments calculated from DFT, panel (b). Remarkably good agreement is seen between the GSE and DFT resulting eigenvectors. Note that the eigenvector provides an additional mode identification mechanism. Experimentally determined modes can be compared with and sorted by calculated modes not only by frequency and amplitude, but also by orientation. Hence, in some instances here, modes observed by GSE and identified by amplitude and direction with a mode calculated by DFT may appear out of frequency sequence, that is, at slightly smaller or slightly larger frequency than predicted by DFT. Thereby, a mode may be found experimentally at a different mode index than predicted by the sequence of DFT calculated frequencies. This occurs here for modes 3 and 4 as well as for modes 17 and 18. The experimental TO mode vector orientations agree within 25$^\circ$ with corresponding calculated modes with the exception of modes 8, 16, 17 and 18. Mode 8 is has the largest disagreement (GSE nearly perpendicular to DFT) which could be explained by its low amplitude and relatively large broadening. It also appears on the shoulder of a much larger nearby mode in GSE data, decreasing sensitivity to mode 8 parameters.

\paragraph{LO mode parameter determination} Cyan solid lines in Figs. \ref{fig:inversetensor} and \ref{fig:tensor} indicate the resulting best match model calculations obtained from Eq. \ref{eq:epsmonoall} using a second independent set of anharmonically broadened Lorentzian oscillators. We find excellent agreement between our wavelength-by-wavelength and model calculated $\varepsilon^{-1}$ and $\varepsilon$. All best match LO model parameters are summarized in Tab. \ref{Table:ExpBu} including amplitude (A$_{\mathrm{LO},l}$), frequency ($\omega_{\mathrm{LO},l}$), broadening ($\gamma_{\mathrm{LO},l}$), anharmonic broadening ($\Gamma_{\mathrm{LO},l}$), and eigenvector direction ($\alpha_{\mathrm{LO},l}$) for all LO modes ($l=1...22$) with $\mathrm{B_u}$ symmetry. Frequencies of the LO modes are indicated by dotted vertical blue lines in Figs. \ref{fig:MM008}, \ref{fig:MM006}, \ref{fig:det}, and \ref{fig:inversetensor} which align with the features observed in the data and the extrema seen in the imaginary part of the inverse dielectric tensor.

\paragraph{LO eigendielectric displacement vectors}

\begin{figure}[hbt]
\centering
\includegraphics[width=\linewidth]{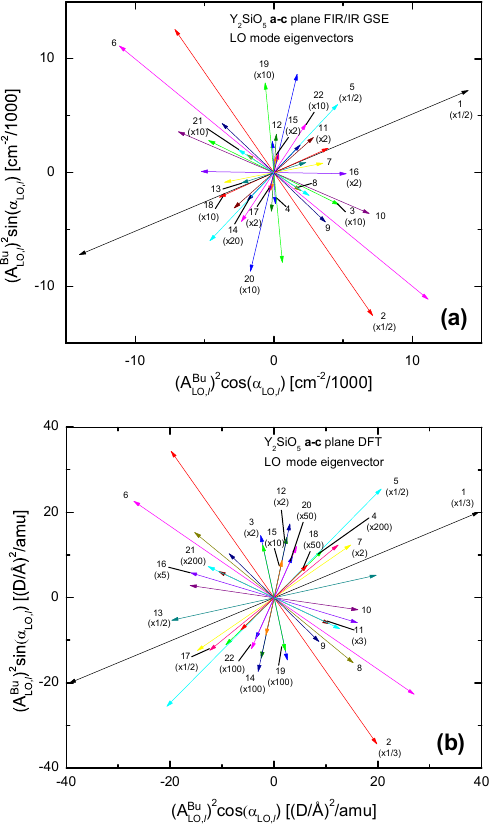}
\caption{\label{fig:LOvectors} (a) Schematic representation of the eigendielectric displacement vectors with GSE analysis determined amplitude $A_{\mathrm{LO},k}^{\mathrm{Bu}}$ and orientation angle $\alpha_{\mathrm{LO},k}$ (with respect to the \textbf{a} axis) of LO modes with $\mathrm{B_u}$ symmetry within the \textbf{a-c} plane. (b) DFT calculated long-wavelength transition dipoles (intensities) of LO modes with $\mathrm{B_u}$ symmetry. }
\end{figure}

Fig. \ref{fig:LOvectors} displays a vector representation of the amplitude and polarization direction parameters (A$^{\mathrm{Bu}}_{\mathrm{LO},k}$ and $\alpha_{\mathrm{LO},k}$) projected onto the \textbf{a-c} plane. Results from the IR/FIR GSE model dielectric function, panel (a), are compared with long-wavelength transition dipole moments calculated from DFT, panel (b). Remarkably good agreement is seen between the GSE and DFT resulting LO eigenvectors. Interestingly, while some LO mode eigenpolarization directions do not deviate very much from their TO counterparts (for example mode 1 with $\alpha_{\mathrm{TO},1}=28.8^\circ$ and $\alpha_{\mathrm{LO},1}=27.2^\circ$), most differ significantly.

\subsubsection{Modes with $\mathrm{A_u}$ symmetry along the \textbf{b} axis}

Resulting mode parameters are described in Tab.~\ref{Table:ExpAu} and dielectric function and inverse dielectric function are given in Figs.~\ref{fig:tensor}(d) and \ref{fig:inversetensor}(d), respectively. Red solid lines show the resulting model calculated dielectric function from Eq.~\ref{eq:epsmonoall} using a set of 23 anharmonic Lorentzian oscillators (Eq.~\ref{eq:AHLO}) for the $\mathrm{A_u}$ symmetry TO modes. Similarly, the solid cyan lines indicate the resulting model calculated dielectric function from Eq. \ref{eq:epsmonoall} using a separate set of 23 anharmonic Lorentzian oscillators (Eq.~\ref{eq:AHLO}) for the $\mathrm{A_u}$ symmetry LO modes. Mode parameters of LO and TO frequencies and broadenings were also determined simultaneously using the BUL form Eq.~\ref{eq:general-eps-broaded} shown in black.

Due to many modes appearing in some narrow frequency regions, sensitivity to separate mode parameters reduces. Contributions from weak modes are easily subsumed by contributions from  strong modes, necessitating several localized spectral best match model analyses (Modes 3, 4, 10, 21 and 23) and some modes required manual setting of parameters (Modes 12, 21, and 23). Frequencies of TO modes with $\mathrm{A_u}$ symmetry are indicated by vertical solid brown lines in Figs.~\ref{fig:MM008}, \ref{fig:MM006}, and \ref{fig:tensor} while frequencies of LO modes with $\mathrm{A_u}$ symmetry are indicated by vertical dotted brown lines in Figs.~\ref{fig:MM008}, \ref{fig:MM006}, and \ref{fig:inversetensor}.

Mode frequencies identified by H\"{o}fer \textit{et al.} are also included in Tab.~\ref{Table:ExpAu} for comparison, where modes 3, 7, 10, 11, 13, 17, 22, and 23 remained unresolved.

\begin{table*}\centering
\caption{\label{Table:ExpAu} Same as for Table \ref{Table:ExpBu} but for phonon mode parameters with $\mathrm{A_u}$ symmetry.}
\begin{ruledtabular}
\begin{tabular}{{l}|{c}{c}{c}{c}{c}{c}{c}{c}|{c}}
Mode & $\omega_{\mathrm{TO}}$(cm$^{-1}$) & $\omega_{\mathrm{LO}}$(cm$^{-1}$)& $\gamma_{\mathrm{TO}}$(cm$^{-1}$) & $\gamma_{\mathrm{LO}}$(cm$^{-1}$)& A$_{\mathrm{TO}}$(cm$^{-1}$) & $\Gamma_{\mathrm{TO}}$(cm$^{-1}$)& A$_{\mathrm{LO}}$(cm$^{-1}$)& $\Gamma_{\mathrm{LO}}$(cm$^{-1}$)&$\bar{v}_j$\cite{Hofer_2016} \\
\hline
1 & 960.0(6) & 988.5(0) & 6.(4) & 9.5(0) & 9(4) & -1.(2) & 238.(1) & -0.(5)&960.6 \\
2 & 914.2(8) & 956.(7) & 5.8(4) & 5.(9) & 43(3) & -8.(9) & 5(7) & -0.(1) &914.4 \\
3 & 887.1(6)$^b$ & 897.5(4)$^b$ & 7.8(9)$^b$ & 4.9(2)$^b$ & 30(9)$^b$ & 9.(9)$^b$ & 38.9(4)$^b$ & 0.18(4)$^b$ & - \\
4 & 883.1(6)$^b$ & 884.8(8)$^b$ & 5.6(0)$^b$ & 6.5(6)$^b$ & 28(4)$^b$ & -0.(1)$^b$ & 8.0(4)$^b$ &  -0.05(5)$^b$ & 884.9 \\
5 & 593.(9) & 620.5(0) & 7.(7) & 14.2(5) & 22(6) & -1(0) & 121.(8) & -1.(6) & 592.8 \\
6 & 560.(1) & 570.(8) & 12.(4) & 12.(6) & 18(9) & -1(6) & 5(1) & 1.(3) & 561.1\\
7 & 54(6) & 54(8) & 1(8) & 1(3) & 1(2)0 & 1(7) & 1(2) & 0.(9) & - \\
8 & 515.(0) & 516.(8) & 6.(8) & 6.(4) & 6(6) & 1.(9) & 2(2) &  0.(0) & 510.9 \\
9 & 439.8(4) & 473.6(5) & 7.(9) & 13.3(2) & 214.(6) & -13.(1) & 10(7) & -1.(3) & 437.1 \\
10 & 411.(9) & 412.(6) & 2(8) & 2(7) & 3(9) & -2.(4) & (7) & 1.(6) & - \\
11 & 40(3) & 410.0(1) & 1(5) & 6.(9) & 1(6) & -0.(1) & 72.(6) & 0.(8) & - \\
12 & 349.6(5) & 40(3) & 8.2(2) & 1(4) & 38(2) & 1(2) & 10$^a$ & -(2) & 345.0 \\
13 & 339.(5) & 341.(7) & 7.(1) & 8.(1) & 19(8) & -2(4) & 6.(8) & -0.3(1) & - \\
14 & 309.9(2) & 317.1(0) & 6.(6) & 5.(7) & 20(8) & (6) & 19.(3) & 0.0(7) & 308.8\\
15 & 270.(1) & 277.7(8) & 3.(6) & 3.(9) & 9(0) & -(3) & 29.(4) & 0.0(5) & 269.9\\
16 & 250.5(2) & 266.(6) & 3.8(2) & 2.(7) & 29(4) & (9) & 16.(0) & -0.0(7) &248.8 \\
17 & 225.(3) & 229.5(3) & 3.(7) & 2.(3) & 1(1)0 & (2)0 & 12.(2) & 0.0(1)&- \\
18 & 223.(5) & 224.(3) & 3.(7) & 2.(3) & 1(2)0 & -(1)0 & 2.(2) & -0.0(2) &223.3 \\
19 & 201.6(2) & 207.(0) & 3.4(5) & 3.(1) & 12(3) & 1(0) & 14.(1) & -0.0(4) &200.9 \\
20 & 188.8(3) & 196.6(8) & 2.3(2) & 3.(8) & 249.(4) & -2(0) & 9.5(2) & -0.04(2)&188.5  \\
21 & 154.(1)$^b$& 154.(5)$^b$ & 2$^a$ & 2$^a$ & 4(3)$^b$ & (9)$^b$ & (3)$^b$ & 0.0(6)$^b$&169.9   \\
22 & 114.(2) & 115.(0) & 2.(1) & 2.(5) & 4(3) & -(3) & 3.(5) & -0.0(2)&- \\
23 & 70$^a$ & 73$^a$ & 5$^a$ & 2$^a$ & 4(6)$^b$ & 0.(8)$^b$ & (3)$^b$ & 0.0(6)$^b$&-  \\

		\end{tabular}
        \end{ruledtabular}
\begin{flushleft}
\footnotesize{$^a${Manually set parameter held constant throughout fitting procedure.}}\\
\footnotesize{$^\textrm{b}${Parameter fit in a local region, held constant in full spectral fit procedure.}}\\
\end{flushleft}
\end{table*}

\subsubsection{TO-LO rule}\label{sec:TOLOrule} The TO-LO rule can best be inspected within the BUL form in Eq.~\ref{eq:general-eps-broaded}. If a switch occurs within a set of ascending frequencies $\omega_{\mathrm{TO},l} < \omega_{\mathrm{LO},l} < \omega_{\mathrm{TO},l+1}< \omega_{\mathrm{LO},l+1} \dots$, for example, $\omega_{\mathrm{TO},l} < \omega_{\mathrm{TO},l+1} < \omega_{\mathrm{LO},l} < \omega_{\mathrm{LO},l+1}\dots$, then the BUL form produces negative imaginary parts in the spectral region between $\omega_{\mathrm{TO},l+1} \dots \omega_{\mathrm{LO},l}$. This is obviously unphysical for a dielectric function, which can be measured along a certain, fixed coordinate direction. However, the determinant function in a low-symmetry material does not represent a directly measurable quantity. Rather, it serves as a spectral indicator for the frequencies of TO and LO modes, as shown in this paper. Furthermore, and accordingly, the determinant produces negative imaginary parts when the order of TO and LO modes within the monoclinic plane is such that the TO-LO rule is broken. Specifically, between $\mathrm{B_u}$ modes 17$\rightarrow$16, 14$\rightarrow$13, 11$\rightarrow$10, 8$\rightarrow$7, 7$\rightarrow$6, 6$\rightarrow$5, 4$\rightarrow$3 and 2$\rightarrow$1. Previously we have observed this TO-LO rule broken for monoclinic $\beta$-Ga$_2$O$_3$\cite{Schubert_2016} but not for monoclinic CdWO$_4$.\cite{Mock_2017} In these cases, when a second TO mode is observed before the next subsequent LO mode, the imaginary part of the determinant is observed to go negative and the imaginary part of the inverse determinant is observed to go positive. Note that the TO-LO rule holds true for all  $\mathrm{A_u}$ modes.

\subsubsection{High frequency and static dielectric constant}

\begin{table}
\caption{\label{tab:constants} Best match model parameters for high frequency dielectric constants along with static dielectric constants extrapolated from the model to $\omega$=0 for each tensor element.}
\begin{ruledtabular}
\begin{tabular}{{l}{c}{c}{c}{c}}
    & $\varepsilon_{\mathrm{xx}}$ & $\varepsilon_{\mathrm{yy}}$ & $\varepsilon_{\mathrm{xy}}$ & $\varepsilon_{\mathrm{zz}}$ \\
  \hline
$\varepsilon_{\infty}$ & 3.16(6) & 3.12(7) & 0.002(7) & 3.11(4) \\
$\varepsilon_{\mathrm{DC}}$ & 10.96(5) & 9.80(1) & 0.12(5) & 11.47(4) \\
\hline
$\varepsilon_{\infty,\mathrm{DFT}}$ & 3.570 & 3.549 & -0.041 & 3.650 \\
\end{tabular}
\end{ruledtabular}
\end{table}

Static and high frequency dielectric constants obtained in this work are summarized in Tab. \ref{tab:constants}. Static dielectric constants $\varepsilon_{\mathrm{DC}}$ for each tensor element were extrapolated from the model calculation at $\omega=0$, and the high frequency dielectric constant was an offset parameter in the best match model analysis. From these, it is determined that the generalized LST relation described by Schubert in Ref.~\onlinecite{Schubert_2016_LST} is well satisfied.

\section{Conclusions}

The frequency dependence of four independent Cartesian tensor elements of the dielectric function for Y$_2$SiO$_5$ were determined using generalized spectroscopic ellipsometry with a dielectric function tensor model approach from 40-1200~cm$^{-1}$. Three different surfaces cut perpendicular to a principle axis were investigated. We match the spectral dependence of the four wavelength-by-wavelength determined dielectric function tensor elements as well as the four inverse tensor elements along with the determinant and its inverse to those rendered by our monoclinic model in order to determine the 22 pairs of transverse and longitudinal optical phonon modes with $\mathrm{B_u}$ symmetry and 23 pairs with $\mathrm{A_u}$ symmetry. We make use of two independent sets of anharmonic oscillators to describe TO and LO mode parameters and their eigendielectric displacement vectors within the \textbf{a-c} plane. We report and compare our experimental findings to density functional theory calculations. We discuss the observation of the violation of the TO-LO rule for polarization within the monoclinic plane. We report the static and high frequency dielectric tensor constants and find that the generalized Lyddane-Sachs-Teller relation is well satisfied for Y$_2$SiO$_5$.

\section{Acknowledgments} The authors thank Larry Alegria (Scientific Materials Corp.) for detailed information on the samples. This work was supported in part by the National Science Foundation (NSF) through the Center for Nanohybrid Functional Materials (EPS-1004094), the Nebraska Materials Research Science and Engineering Center (MRSEC) (DMR-1420645) and awards CMMI 1337856 and EAR 1521428. The authors further acknowledge financial support by the University of Nebraska-Lincoln, the J.~A.~Woollam Co., Inc., and the J.~A.~Woollam Foundation. The DFT calculations were performed using the computing resources of the Holland Computing Center and the Center for Nanohybrid Functional Materials at the University of Nebraska-Lincoln.
\end{document}